%% file: main.tex
\documentclass[lettersize,journal]{IEEEtran}

\input{settings.tex}

\input{terms.tex}

\begin{document}

\title{Model Sharing Bayesian Optimization\\for Mechanical Wi-Fi Antenna Control in Multi-Station Environments}

\author{Akihito~Taya,
        and~Yuuki~Nishiyama,
        and~Kaoru~Sezaki,~\IEEEmembership{Member,~IEEE}
\thanks{A. Taya
is with Institute of Industrial Science, The University of Tokyo, Meguro-ku, Tokyo 153-8505, Japan.\protect\\
E-mail: taya-a@iis.u-tokyo.ac.jp}%
\thanks{Y. Nishiyama and K. Sezaki
are with Center for Spatial Information Science, The University of Tokyo, Kashiwa-shi, Chiba 277-8568, Japan.}
\thanks{Manuscript received April 19, 2021; revised August 16, 2021.}%
\thanks{This work was supported by JSPS KAKENHI Grant Number JP23K26109, JP24K20759 and the Telecommunications Advancement Foundation.}%
}


\markboth{Journal of \LaTeX\ Class Files,~Vol.~14, No.~8, August~2021}%
{Shell \MakeLowercase{\textit{et al.}}: A Sample Article Using IEEEtran.cls for IEEE Journals}


\maketitle

\begin{abstract}
Some commercial \WiFi access points used in homes and offices have antennas whose orientations can be adjusted manually.
However, selecting suitable orientations is difficult for non-expert users, particularly in indoor multipath environments where reflections and polarization effects complicate the relationship between antenna orientation and link quality.
We investigate automatic mechanical antenna orientation control for uplink \WiFi communication when the set of active stations changes over time.
Building on our previously developed antenna-control system, we propose a model-sharing \gls{BO} framework that learns the log-capacity of each link using a separate \gls{GPR} model.
The framework reuses these models across active-\gls{STA} combinations and optimizes proportional fairness, equivalently maximizing the geometric mean of the active links' capacities.
Exploiting the linearity of this objective in log-capacity, we also derive a \gls{UCB} acquisition function and its gradient.
We evaluate the framework through simulations driven by measured channel capacities and experiments with a mechanical \WiFi antenna prototype.
The results demonstrate that the proposed model-sharing \gls{BO} achieves better optimization performance than independent state-specific optimization by reusing the models trained on previously observed links.
\end{abstract}


\glsunset{IoT}
\section{Introduction}
\IEEEPARstart{I}{nternet} of Things (\gls{IoT}) networks increasingly use \WiFi to carry uplink data from cameras, \gls{LiDAR} sensors, and other connected devices.
As these devices start and stop their data streams independently, the set of active \glspl{STA} associated with a common \gls{AP} changes over time.
Maintaining fair link performance across such changes is challenging even when the stations remain stationary.

Indoor measurements have shown that antenna tilt and rotation can affect received signal strength through changes in radiation patterns and polarization alignment \cite{hilton2011evaluating}.
Thus, antenna orientation is a physical parameter that may be optimized alongside software-configurable radio settings.
However, antennas on commercial \glspl{AP} typically have fixed or manually adjustable orientations, and selecting favorable angles is difficult for non-expert users in multipath-rich indoor environments.
Software-controlled mechanical actuation offers a way to tune antenna orientation based on observed link quality without requiring an explicit propagation model.

Recent research on \glspl{MA} and \glspl{RA} has explored physical antenna reconfiguration to improve wireless communication performance \cite{zhu2024modeling,zhu2024movable,zheng2025rotatable,zheng2026rotatable,ning2025movable}.
Many studies consider array-element reconfiguration or model-based optimization, although whole-antenna rotation and polarization-aware approaches have also been investigated.
Mechanically actuated systems such as NetBeam \cite{bocanegra2019netbeam} have demonstrated orientation control but typically rely on tightly managed communication architectures.
Our focus instead is the orientation of broad-beam antennas in conventional indoor \WiFi networks with time-varying sets of active \glspl{STA}.

\begin{figure}[!t]
  \centering
  \subfloat{\includegraphics[width=\subfigWidthPhoto,trim=0 0 50 25,clip]{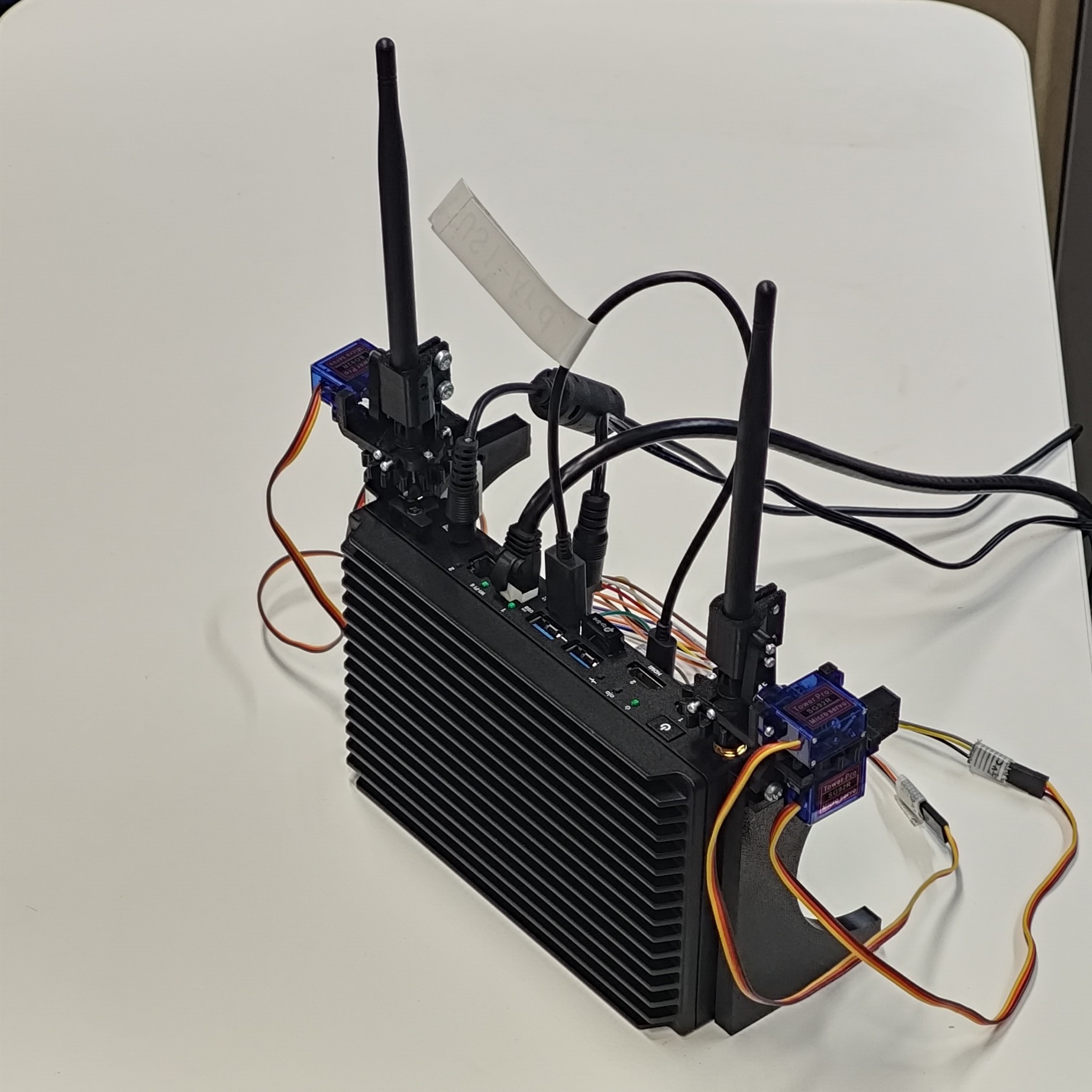}}\hspace{0.5em}
  \subfloat{\includegraphics[width=\subfigWidthPhoto,trim=0 0 50 25,clip]{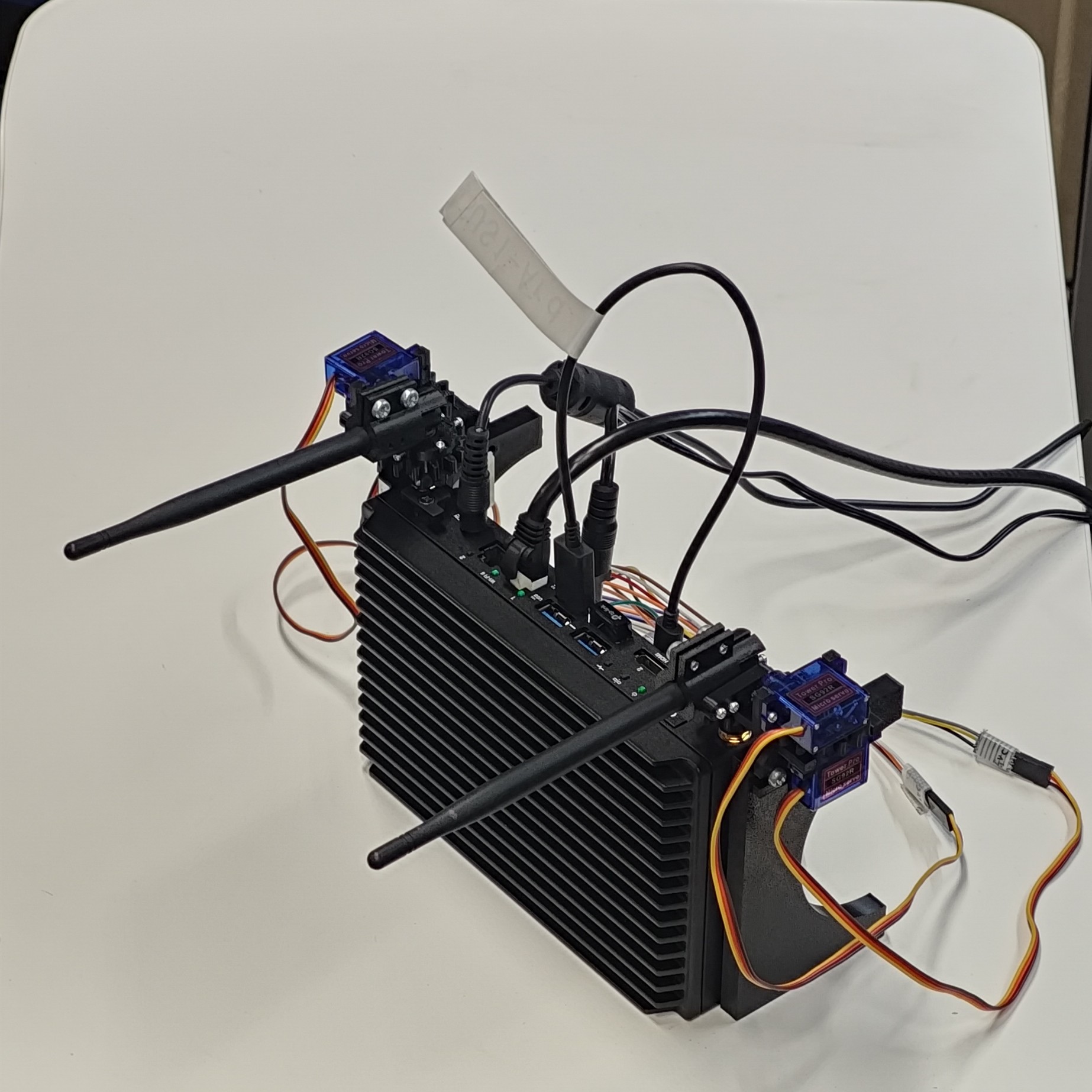}}\\
  \vspace{-0.5em}
  \subfloat{\includegraphics[width=\subfigWidthPhoto,trim=0 0 50 25,clip]{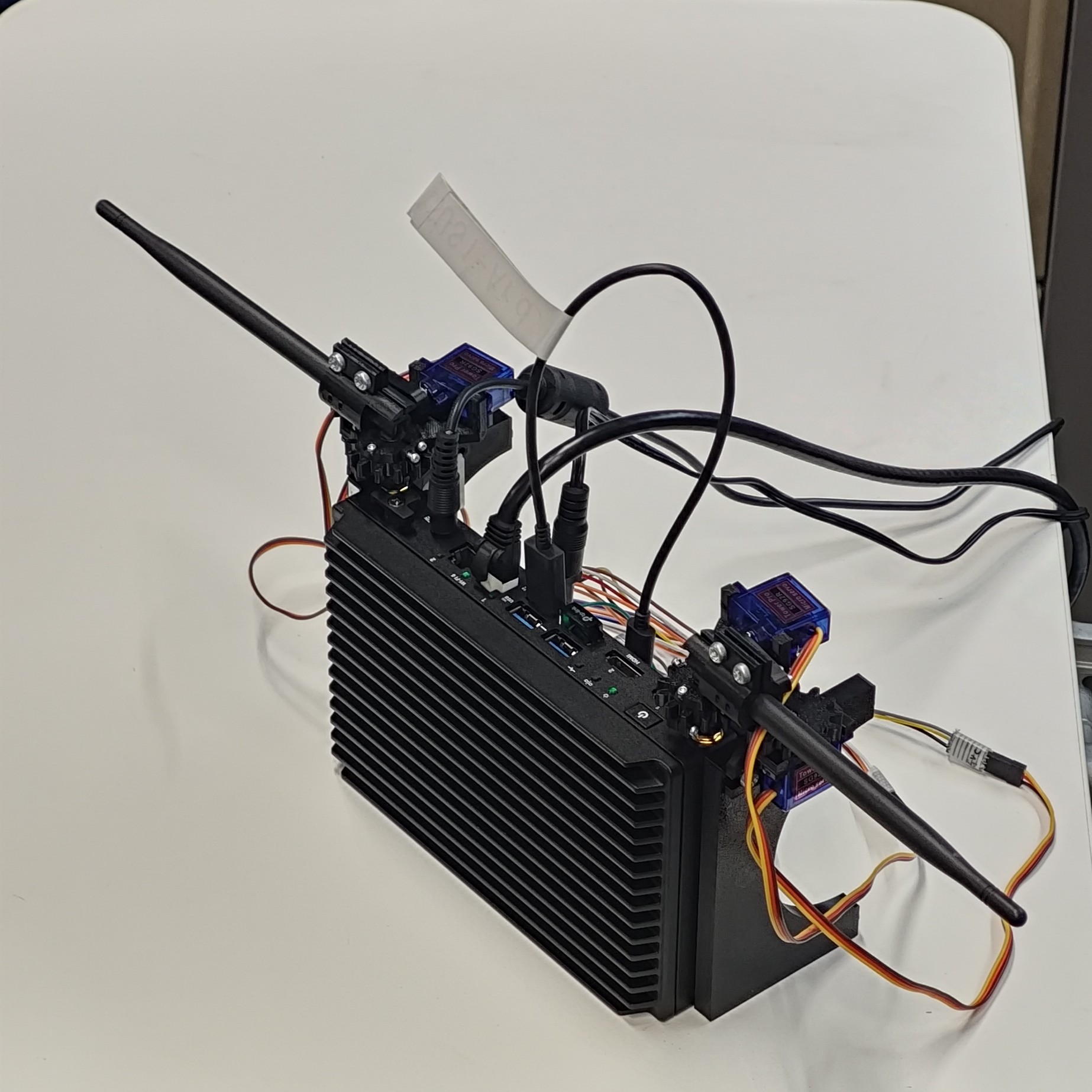}}\hspace{0.5em}
  \subfloat{\includegraphics[width=\subfigWidthPhoto,trim=0 0 50 25,clip]{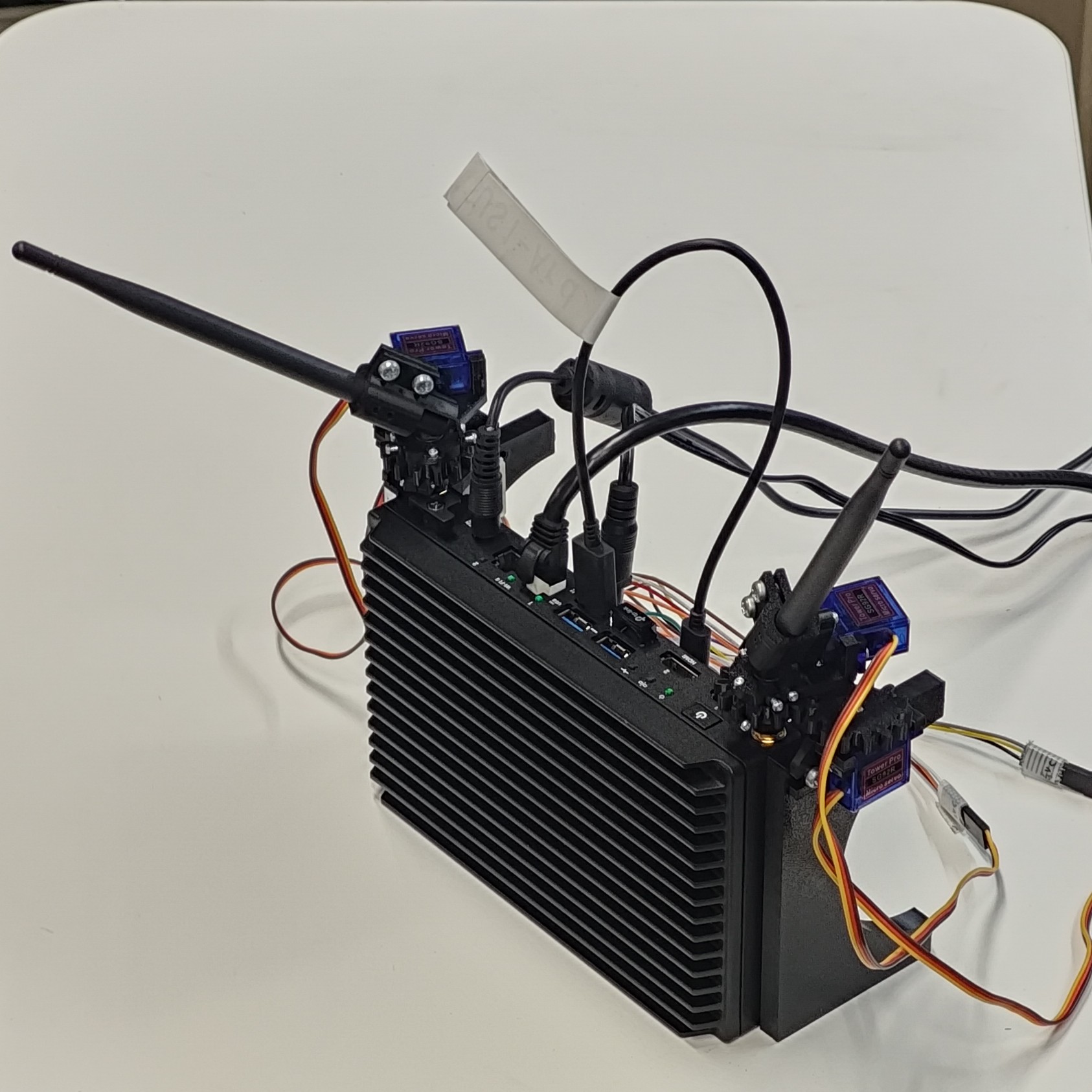}}
  \caption{Mechanical \WiFi antenna system with two independently controllable antennas. The device automatically adjusts the 3D orientation of each antenna to optimize channel capacity.}
  \label{fig:device}
\end{figure}

The orientation that maximizes the capacity of one \gls{STA} need not maximize proportional fairness across multiple active \glspl{STA}.
Consequently, the optimization objective changes as stations become active or inactive, even when the propagation conditions of individual links remain approximately stationary.
Moreover, mechanical actuation is slow relative to wireless frame reception, so antenna control and capacity measurements must operate asynchronously.
Practical systems must also handle occasional persistent changes in propagation conditions; continuously mobile stations and rapidly time-varying channels, however, are outside the scope of this study.

In our previous work, we developed the mechanical \WiFi antenna system shown in Fig.~\ref{fig:device} \cite{taya2026ccnc,taya2026globecom}.
Our initial study demonstrated the feasibility of automatic antenna orientation tuning for improving link capacity \cite{taya2026ccnc}.
We subsequently introduced an asynchronous optimization framework that maintains independent optimizers for different active-\gls{STA} combinations and selects the appropriate optimizer as the active set changes \cite{taya2026globecom}.
The conference framework also incorporated capacity-based detection of persistent environmental changes and optimizer reset, which we retain in this work.

However, maintaining a separate \gls{BO} model for each active-\gls{STA} combination can require up to $2^N - 1$ surrogate models for $N$ \glspl{STA}.
Moreover, observations collected for one combination do not update the models of other combinations, even when they share the same links.
This limits knowledge transfer across active-set changes and can lead to repeated exploration of previously observed link characteristics.

To address this limitation, we exploit the fact that, under approximately stationary and interference-negligible conditions, the channel capacity of a given link can be modeled as a function of antenna orientation independently of the activity of other stations.
Inspired by composite \gls{BO} \cite{astudillo2019bayesian,jain2023bayesian}, we maintain a separate \gls{GPR} model for each link and update only the models of currently active links.
For each active-\gls{STA} combination, the optimization objective is the mean log-capacity of its links, which is equivalent to maximizing their geometric-mean capacity.
Previously learned link models can therefore be reused when a different active set is encountered, without learning a new surrogate for the entire combination.
Assuming independent link-wise \gls{GPR} predictive distributions, we further derive a \gls{UCB} acquisition function and its gradient for this proportional-fairness objective.

The contributions of this paper are summarized as follows:%
\footnote{
An earlier version of this work was presented at IEEE GLOBECOM 2026 \cite{taya2026globecom}.
The asynchronous architecture, state-specific optimization, and environment-change detection and reset mechanism were introduced in that version; the present paper extends it with per-link model sharing and a derived acquisition function.
}
\begin{itemize}
  \item We propose a model-sharing \gls{BO} framework that updates per-link \gls{GPR} models using observations from active \glspl{STA} and reuses them across different active sets, reducing the number of surrogate models from up to $2^N - 1$ to $N$.
  \item We analytically derive a \gls{UCB} acquisition function and its gradient for proportional fairness by exploiting the linearity of the objective in log-capacity.
  \item We conduct real-value-based simulations to statistically compare the performance of optimization methods across multiple active-set sequences.
  \item We evaluate the proposed framework through real-world experiments using our mechanical \WiFi antenna prototype under controlled changes in station activity, demonstrating the benefits of reusing link models when a new active-\gls{STA} combination is encountered.
\end{itemize}

\section{Related Work}
Beamforming is a conventional approach to controlling radio propagation by steering antenna directivity mechanically or electronically \cite{uchendu2016survey,abbasi2019millimeter,kutty2016beamforming}.
Pal \etal \cite{pal2018conformal} proposed a conformal beam-steering antenna electronically controlled by a Raspberry Pi to switch beam directions and demonstrated improved throughput in indoor environments.
Such beam-steering techniques generally aim to provide directional links.
In contrast, our mechanical system controls the 3D orientations of broad-beam antennas (e.g., omnidirectional antennas) to serve multiple devices through a common \gls{AP}.
The orientation control also changes antenna polarization, which can affect channel capacity, as demonstrated in \cite{taya2026ccnc}.

\Gls{MA} systems can be classified by their actuation scale into element-level control of individual radiating elements and array-level control of an antenna structure as a whole \cite{ning2025movable}.
Recent studies on element-level \glspl{MA} \cite{zhu2024modeling,zhu2024movable} have demonstrated communication performance improvements through antenna position adjustment.
These systems adjust individual antenna elements to improve channel conditions and beamforming performance.
In particular, \glspl{RA} \cite{zheng2025rotatable,zheng2026rotatable} have been studied for improving communication performance by adjusting the orientations of antenna elements.

In addition to element-level control, whole-array control and the reorientation of stand-alone antennas have also been studied.
Mechanical antenna tilting has long been studied for cellular base stations to improve coverage and capacity \cite{Niemela2004impact,athley2010impact}.
For 3D orientation control, NetBeam \cite{bocanegra2019netbeam} investigated distributed beamforming using multiple transmitters equipped with servo motors.
Although NetBeam provides insights from indoor experiments, it relies on a centrally managed architecture using \glspl{SDR} for tight synchronization and \gls{CSI} feedback.
Recent six-dimensional \gls{MA} studies have jointly optimized antenna position and orientation \cite{shao2025modeling,shao2025movable,pi2026movable}.
Shao \etal \cite{shao2025modeling} developed a channel model and a continuous optimization method for six-dimensional antenna adjustment.
Their subsequent work \cite{shao2025movable} proposed methods for cases with known or unknown channel statistics and considered discrete antenna adjustments compatible with stepper-motor control.
Both studies evaluated their methods through simulations.
Pi \etal \cite{pi2026movable} extended joint position-and-orientation optimization to a centrally coordinated multi-\gls{AP} architecture rather than a conventional \WiFi network.
These studies use geometry-based channel models to predict how antenna movement and rotation affect channel conditions.
Wei \etal \cite{wei2026polarization} developed a polarization- and mutual-coupling-aware model for \glspl{RA} and evaluated alternating optimization using a 3GPP channel model and experimental measurements.

In contrast, we consider broad-beam antennas in indoor environments such as homes and offices.
We use \gls{GPR} to learn how channel capacity varies with antenna orientation and predict capacities at unmeasured orientations without requiring an explicit geometry-based propagation model.
Our focus is a model-sharing \gls{BO} framework that learns separate link-capacity functions and optimizes proportional fairness as the active-\gls{STA} set changes while individual link conditions remain approximately stationary.

\section{Preliminary: Bayesian Optimization} \label{eq:preliminary}
\subsection{Basics}
\Gls{BO} \cite{garnett2023bayesian,shahriari2015taking} is a black-box optimization method that iteratively selects the next evaluation point based on a surrogate model of the objective function.
Let $f:\mathcal{X}\to\RealMat$ be a black-box function to be maximized, where $\mathcal{X}\subseteq\RealMat^d$ is the input space.
\Gls{BO} constructs a surrogate model of $f$ using \gls{GPR}, which provides a predictive distribution for the function value at any input $x\in\mathcal{X}$.
Acquisition functions are then used to balance exploration and exploitation when selecting the next evaluation point.
A commonly used acquisition function is the \gls{UCB} criterion, defined as follows:
\begin{align}
  \Acq(x) = \meanEst(x) + \ucbCoef \stdEst(x), \label{eq:basic_ucb}
\end{align}
where $\meanEst(x)$ and $\stdEst(x)$ are the mean and standard deviation of the surrogate model at point $x$, respectively, and $\ucbCoef$ is a hyperparameter that controls the exploration--exploitation trade-off.
Because $\stdEst(x)$ tends to decrease near observed inputs and remain higher in unexplored regions, \eqref{eq:basic_ucb} balances exploitation of promising regions with exploration of uncertain ones.

Concretely, \gls{BO} iteratively selects the next evaluation point $x^*$ as the one that maximizes the acquisition function; upon evaluating $f(x^*)$, the surrogate model is updated with the new observation, and the process repeats until a stopping criterion is met.
For continuous search spaces, differentiable acquisition functions can be optimized using gradient-based methods such as \gls{L-BFGS-B} \cite{zhu1997algorithm,georgiou2025deterministic}.

Unlike population-based methods such as particle swarm optimization and genetic algorithms, \gls{BO} is designed to select informative points when objective evaluations are expensive.
In our antenna-orientation optimization system, evaluations cannot be performed in parallel, and each evaluation takes several seconds because of antenna movement and subsequent capacity measurement.
Therefore, \gls{BO} is well suited to this sequential, measurement-intensive optimization problem.

\subsection{Composite Bayesian Optimization}
Astudillo and Frazier \cite{astudillo2019bayesian} proposed a composite \gls{BO} framework that optimizes a function composition $f(g(x))$ by modeling the unknown inner function using a multi-output \gls{GPR} while exploiting the known outer function.
The authors then extended this framework to handle networks of intermediate functions \cite{astudillo2021bayesian}.
Multiple-input composite \gls{BO} was also studied in \cite{jain2023bayesian}, which introduced the acquisition function referred to as \gls{cUCB}.
Related extensions include optimization with multiple components and target values \cite{jeong2021bayesian} and cost-aware partial evaluation of function networks \cite{buathong2024bayesian}.
Huang \etal \cite{huang2026combinatorial} studied tensor-output \gls{BO}, including the selection of subsets of outputs contributing to the objective.

Because antenna-orientation optimization for multiple active \glspl{STA} can be formulated as a composite optimization problem, we draw on composite \gls{BO} methods.
In our problem, the active set of \glspl{STA} changes exogenously over time, which alters both the available component observations and the composition defining the optimization objective.
Thus, different active sets yield different compositions of the same underlying link functions, requiring the acquisition function and observed components to be adapted to the current set.
Reusing surrogate models of these link functions across active sets can improve sample efficiency.

\section{System Model}

We consider a set of \glspl{STA} $\Stations$ that independently start and stop uplink data streams to a common \gls{AP}, thereby changing the active-\gls{STA} set over time.
A typical use case is an indoor \gls{IoT} edge network in which an \gls{AP} receives high-bandwidth uplink sensor streams, e.g., data from cameras, LiDAR sensors, or robots, from multiple sensing devices.
Although we focus on the uplink, the framework could also be extended to downlink scenarios if suitable per-link channel measurements are available.

Unlike conventional \glspl{AP} with fixed antenna orientations, our proposed \gls{AP} adjusts its antenna orientations $\AngleSet$ to optimize channel capacity.
Specifically, the device independently controls two antennas in both elevation $\theta_j$ and azimuth $\varphi_j$ (for antenna $j\in\{0,1\}$), yielding a four-degree-of-freedom configuration space: $\AngleSet=(\theta_0, \varphi_0, \theta_1, \varphi_1)$.

Upon receiving data frames from \gls{STA} $i\in\Stations$ at a given orientation $\AngleSet$, the \gls{AP} measures the \gls{CSI} $\CSI_{i,\SCindex}(\AngleSet)\in\ComplexMat^{\NumRxAntenna \times \NumTxAntenna}$, where $\SCindex$, $\NumRxAntenna$ and $\NumTxAntenna$ denote the subcarrier index, the number of receive antennas, and the number of transmit antennas, respectively.
Using this \gls{CSI}, the \gls{AP} computes the achievable channel capacity, which serves as a physical-layer performance metric, and optimizes $\AngleSet$ to maximize proportional fairness among all active \glspl{STA}.

The \gls{MIMO}-\gls{OFDM} channel capacity $\Capacity_i(\AngleSet)$ between \gls{STA} $i\in\Stations$ and the \gls{AP} is calculated as \cite{goldsmith2003capacity}:
\begin{align}
  \Capacity_i (\AngleSet)
    = \BWperSubCarrier \sum_{\SCindex=1}^{\NumSubcarrier} \log_2
      \[ 
        \det \(\IdentityMat + \frac{1}{\NoiseDensity \BWperSubCarrier} \CSI_{i,\SCindex}(\AngleSet) \CSI_{i,\SCindex}(\AngleSet)^\Hermit\) 
      \], \label{eq:capacity}
\end{align}
where $\BWperSubCarrier$, $\NumSubcarrier$, and $\NoiseDensity$ denote the subcarrier bandwidth, the number of subcarriers, and the noise density per antenna, respectively.
Additionally, $\IdentityMat$ is the $\NumRxAntenna \times \NumRxAntenna$ identity matrix and $[\cdot]^\Hermit$ denotes the Hermitian transpose.
Note that $\CSI_{i,\SCindex}(\AngleSet)$ is treated as an effective channel matrix that includes the received signal power scaling.

To balance overall channel capacity and promote fairness among users, we adopt proportional fairness \cite{shi2014fairness} based on the time-averaged capacity $\AvgCapacity_i(\AngleSet)$ over a window $\AvgTime$ as our optimization objective.
First, the geometric mean of the channel capacity of the active \glspl{STA} is computed as:
\begin{align}
  \GeoMCapacity_{\StaSet}(\AngleSet) \defeq \(\prod_{i \in \StaSet} \AvgCapacity_i (\AngleSet)\)^{1/\Card{\StaSet}}, \label{eq:geomean}
\end{align}
where $\Card{\StaSet}$ denotes the number of active \glspl{STA}.
We define the proportional fairness metric $\Fairness_{\StaSet}(\AngleSet)$ for the active stations $\StaSet\subseteq\Stations$ as the logarithm of the geometric mean:
\begin{align}
  \Fairness_{\StaSet}(\AngleSet)
    \defeq& \log \GeoMCapacity_{\StaSet}(\AngleSet) \nonumber \\
    =& \frac{1}{\Card{\StaSet}} \sum_{i \in \StaSet} \log \(\AvgCapacity_i (\AngleSet)\). \label{eq:fairness}
\end{align}
Since the ideal antenna orientation inherently depends on which devices are active, the system aims to find the optimal configuration $\AngleSetIdeal_\StaSet \defeq \argmax_{\AngleSet} \Fairness_{\StaSet}(\AngleSet)$ for each unique \gls{STA} combination $\StaSet$.

Practically, commercial hardware often reports a normalized \gls{CSI}, $\NormCSI_{i,\SCindex}(\AngleSet) = \gainCoef \CSI_{i,\SCindex}(\AngleSet)$.
Although the scaling factor $\gainCoef$ is typically undisclosed, it can be derived from the \gls{RSSI}.
Because \gls{RSSI} encapsulates the total signal and noise power across all subcarriers, it satisfies:
\begin{align}
  \rssi = \sum_{\SCindex=1}^{\NumSubcarrier} \|\CSI_{i,\SCindex}(\AngleSet)\|^2 + \NumSubcarrier \NumRxAntenna \NoiseDensity \BWperSubCarrier,
\end{align}
where $\|\cdot\|$ denotes the Frobenius norm.
Consequently, $\gainCoef$ is extracted as:
\begin{align}
  \gainCoef^2 = \frac{\sum_{\SCindex=1}^{\NumSubcarrier} \|\NormCSI_{i,\SCindex}(\AngleSet)\|^2}{\rssi - \NumSubcarrier \NumRxAntenna \NoiseDensity \BWperSubCarrier}.
\end{align}
Substituting this into \eqref{eq:capacity} yields the practical capacity formula:
\begin{align}
  \Capacity_i (\AngleSet)
    &= \BWperSubCarrier \nonumber \\
    &\times \sum_{\SCindex=1}^{\NumSubcarrier} \log_2
      \[ 
        \det \(\IdentityMat + \frac{1}{\gainCoef^2 \NoiseDensity \BWperSubCarrier} \NormCSI_{i,\SCindex}(\AngleSet) \NormCSI_{i,\SCindex}(\AngleSet)^\Hermit\) 
      \]. \label{eq:capacity-measurement}
\end{align}

In our system, the active or inactive state of each \gls{STA} is determined by the presence or absence of its uplink frames.
In the experimental evaluation, a \gls{STA} is considered inactive if no frames are received from it for a specified period.

\section{Mechanical Antenna Control System for Multi-Station Optimization}
\subsection{Asynchronous System Architecture} \label{sec:async_architecture}
\begin{figure}[!t]
  \centering
  \subfloat[Sequence diagram of the antenna tuning framework]{
    \resizebox{!}{180pt}{\input{images/sequence.tex}}\label{fig:sequence_diagram_sub}
  }\\
  \subfloat[State-specific BO]{
    \resizebox{100pt}{!}{\input{images/framework_ss.tex}}\label{fig:framework_ss}
  }
  \subfloat[Model-sharing BO]{
    \resizebox{130pt}{!}{\input{images/framework_bo.tex}}\label{fig:framework_bo}
  }
  \caption{
    Sequence diagram of the antenna tuning framework and the developed optimization frameworks.
    While the state-specific optimization framework maintains independent optimizers for each active-\gls{STA} combination, the proposed model-sharing framework shares per-link GPR models across different combinations.
  }
  \label{fig:sequence_diagram}
\end{figure}
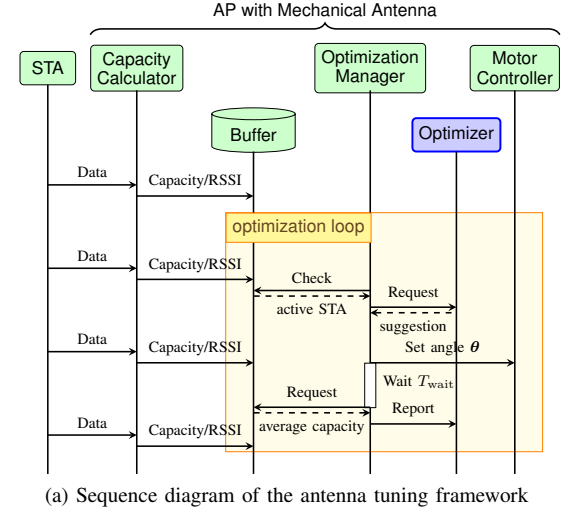
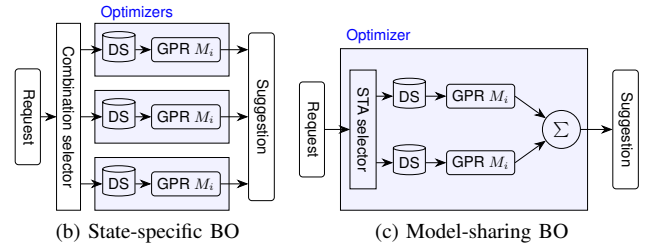

In an unmanaged network such as \WiFi, an \gls{AP} cannot prescribe the exact packet transmission times of individual \glspl{STA}.
In addition, mechanical antenna movement takes tens of milliseconds, whereas frame reception and capacity estimation occur on shorter timescales.
We therefore employ the asynchronous architecture developed in our conference work \cite{taya2026globecom}, illustrated in Fig.~\ref{fig:sequence_diagram}.
A Capacity Calculator continuously processes received frames and maintains a moving average of the channel capacity for each observable link in a shared buffer.
Independently, the Optimization Manager requests a suggested antenna orientation from the optimizer, drives the actuator, and reads the capacities from the buffer for the objective evaluation.
A stabilization interval $\WaitTime$ is imposed after each movement so that measurements taken at the preceding orientation are not used as observations for the new one.
An evaluation is discarded if a required link has no valid post-movement measurement.
This decoupling avoids blocking the capacity calculation while the antenna is moving, without assuming that packet arrivals are synchronized with antenna control.

\subsection{Concurrent Optimization for Multiple Active Station Combinations} \label{sec:method_globecom}

We first summarize the state-specific optimization framework introduced in \cite{taya2026globecom}, which forms the basis of our extension.
As proportional fairness may be non-convex with respect to the multiple antenna angles, we treat the objective $\Fairness_{\StaSet}(\AngleSet)$ as a black-box function.
\Gls{BO} \cite{garnett2023bayesian} is suitable for this setting because each physical evaluation incurs antenna movement and measurement delay.
In the conventional implementation, an independent optimizer $\Opt{\StaSet}$ models the scalar objective for each nonempty active-\gls{STA} set $\StaSet$; its next orientation is selected using the standard \gls{UCB} acquisition function \eqref{eq:basic_ucb}.

\begin{figure}[!t]
  \begin{algorithm}[H]
    \caption{State-specific optimization framework from \cite{taya2026globecom}} \label{alg:globecom}
    \begin{algorithmic}[1]
      \While{optimization is running}
        \State Obtain the current active \gls{STA} set $\StaSet$
        \OnelineIf{$\StaSet = \emptyset$} Wait $\WaitTime$ and \textbf{continue}
        \State Select the state-specific optimizer $\Opt{\StaSet}$
        \State $\AngleSet \gets \OptSuggest$ \AlgComment{best known if converged}
        \State Move antennas to $\AngleSet$ and wait $\WaitTime$
        \OnelineIf{a required link measurement is invalid} \textbf{continue}
        \State Obtain $\AvgCapacity_i(\AngleSet)$ for all $i\in\StaSet$
        \If{$\OptConv$}
          \If{a persistent capacity change is detected}
            \State $\OptReset$
          \EndIf
        \Else
          \State $\OptReport{\AngleSet}{\Fairness_\StaSet(\AngleSet)}$
          \State Update the state-specific best orientation and baseline
        \EndIf
      \EndWhile
    \end{algorithmic}
  \end{algorithm}
\end{figure}

As shown in Algorithm~\ref{alg:globecom}, only the optimizer associated with the current active set is used; all other optimizers retain their learned states until their corresponding combinations reappear.
If any \gls{STA} required for the pending measurement becomes inactive during actuation, the evaluation is skipped rather than assigned to the wrong active set.

After an optimizer reaches its stopping criterion, the system revisits its best-known orientation $\AngleSetBest_\StaSet$ and compares the observed per-link capacities with the stored baselines $\CapBest_{\StaSet,i}$.
Specifically, the relative deviation for $i\in\StaSet$ is
\begin{align}
  \gap_i &= \frac{
      \left|\AvgCapacity_i(\AngleSetBest_\StaSet)-\CapBest_{\StaSet,i}\right|
    }{
      \CapBest_{\StaSet,i}
    }. \label{eq:capacity_change}
\end{align}
An environmental change is declared if the deviation exceeds $\threshold$ for at least one active link during $\MaxChangeAccept$ consecutive evaluations.
The corresponding optimizer is then reset for re-optimization.
This detection mechanism is inherited from \cite{taya2026globecom}.

\subsection{Model-Sharing Bayesian Optimization} \label{sec:method_extension}
\subsubsection*{Motivation and relation to composite optimization}
The state-specific approach may maintain up to $2^{\Card{\Stations}}-1$ independent \gls{GPR} models, one for each nonempty combination of active \glspl{STA}.
More importantly, an observation from one combination cannot directly update the surrogate model for another combination, even if they contain a common link.
Composite \gls{BO} addresses a related modeling problem by constructing surrogate models for constituent functions \cite{astudillo2019bayesian,jain2023bayesian}.
In particular, cUCB \cite{jain2023bayesian} combines the individual upper-confidence estimates through a known, generally nonlinear composition function.
Utilizing this concept, we developed an antenna control framework in which both the observed links and the optimization objective change with the active-\gls{STA} set.
This approach, which shares the underlying channel capacity across different combinations, allows the newly encountered active combination to reuse previously learned per-link models, enabling efficient optimization for the considered problem where the objective functions grow exponentially with the number of \glspl{STA}.
We further exploit the linear structure of \eqref{eq:fairness} to derive the predictive uncertainty of the objective directly.

\subsubsection*{Active-link observations and updating shared models}
Let $\StaSet_t\subseteq\Stations$ be the set of active \glspl{STA} at optimization round $t$, and define the per-link response $\logCap_i(\AngleSet)=\log\AvgCapacity_i(\AngleSet)$.
We assume that, within a period for which the propagation environment is approximately stationary, this channel-capacity metric for link $i$ depends on the antenna orientation but not on the identities of other simultaneously active \glspl{STA}.
This is appropriate for the \gls{CSI}-derived physical-layer link-capacity metric considered here; it does not imply that medium-access throughput or interference-limited rates are independent of the active set.
The active set determines which link capacities can be observed and which terms enter the proportional fairness objective.

For each \gls{STA} $i$, the proposed optimizer $\OptM$ maintains a dataset $\localDS_i$ and a \gls{GPR} surrogate model $\surrogate{i}$ for $\logCap_i$.
When a valid observation at $\AngleSet_t$ becomes available, the datasets of the active links receive new samples as $\localDS_i \gets \localDS_i \cup \{(\AngleSet_t,\log\AvgCapacity_i(\AngleSet_t))\}$ and the datasets for inactive links remain unchanged.
Then, the surrogate models corresponding to the active links are trained by \gls{GPR} using the updated datasets.

\subsubsection*{Analytically derived acquisition function}
We model each log-capacity function $\logCap_i(\AngleSet)$ by a Gaussian process with predictive mean $\meanEst_i(\AngleSet)$ and standard deviation $\stdEst_i(\AngleSet)$.
We assume that the posterior predictive distributions of the individual link-wise Gaussian process models are independent.
Under this assumption, their linear combination in \eqref{eq:fairness} is Gaussian, and the predictive mean and standard deviation of the proportional fairness objective are
\begin{align}
  \meanEst_\StaSet(\AngleSet)
    &= \frac{1}{\Card{\StaSet}} \sum_{i \in \StaSet} \meanEst_i(\AngleSet), \label{eq:acquisition_inner1} \\
  \stdEst_\StaSet(\AngleSet)
    &= \frac{1}{\Card{\StaSet}} \sqrt{\sum_{i \in \StaSet} \stdEst_i^2(\AngleSet)}. \label{eq:acquisition_inner2}
\end{align}
The resulting \gls{UCB} acquisition function can be expressed as
\begin{align}
  \Acq(\AngleSet;\StaSet)
    &= \meanEst_\StaSet(\AngleSet) + \ucbCoef \stdEst_\StaSet(\AngleSet).
    \label{eq:acquisition}
\end{align}
We refer to \eqref{eq:acquisition} as \gls{propUCB}.
For comparison, applying the \gls{cUCB} construction \cite{jain2023bayesian} to the same linear objective yields $\Card{\StaSet}^{-1}\sum_{i\in\StaSet}\left(\meanEst_i(\AngleSet)+\beta\stdEst_i(\AngleSet)\right)$, where $\beta$ is its exploration parameter.
Both constructions employ the same per-link regression models.
However, they differ in whether standard deviations are summed or the standard deviation of the aggregate objective is calculated.
We compare these acquisition functions within the same model-sharing framework in Sec.~\ref{sec:evaluation}.

The gradient of the proposed acquisition function, which can be used in gradient-based optimization, e.g., \gls{L-BFGS-B}, can be obtained analytically from those of the constituent Gaussian process predictions:
\begin{align}
  \nabla_{\AngleSet} \Acq(\AngleSet; \StaSet)
  &= \frac{1}{\Card{\StaSet}}
     \sum_{i \in \StaSet} \nabla_{\AngleSet} \meanEst_i(\AngleSet) \nonumber\\
  &\quad
   + \frac{\ucbCoef}{\Card{\StaSet}}
     \frac{\sum_{i \in \StaSet} \stdEst_i(\AngleSet) \nabla_{\AngleSet} \stdEst_i(\AngleSet)}
          {\sqrt{\sum_{i \in \StaSet} \stdEst_i^2(\AngleSet)}}.
   \label{eq:acquisition_gradient}
\end{align}
This expression enables gradient-based acquisition maximization using gradients of the individual \gls{GPR} predictions.

\subsubsection*{State management and reset handling}
Although each surrogate model corresponds to an individual \gls{STA}, the optimization objective depends on the current set $\StaSet$.
The optimizer therefore maintains the best proportional fairness score $\FairnessBest_\StaSet$, its corresponding orientation $\AngleSetBest_\StaSet$, and stopping status for each active set, separately from the shared models.
The Optimization Manager separately stores the per-link capacities $\CapBest_{\StaSet,i}$ measured at $\AngleSetBest_\StaSet$ for environment change detection.
While the counter $\Counter_i$ tracks the number of observations for each link to determine whether link $i$ has collected at least $\InitialPoint$ initial observations, the stopping criterion including the maximum number of evaluations is evaluated for each active set $\StaSet$.
Upon reaching the stopping criterion for a particular set, its stored best orientation is used to monitor the corresponding capacities as in Section~\ref{sec:method_globecom}.

We apply the environment-change detection and reset mechanism inherited from \cite{taya2026globecom} to the shared-model framework.
When a persistent change is detected, the models of the currently active links are reset, and the optimization states of all active-\gls{STA} combinations depending on those models are invalidated.

\begin{figure}[!t]
  \begin{algorithm}[H]
    \caption{Dynamic model-sharing \gls{BO}} \label{alg:proposed}
    \begin{algorithmic}[1]
      \State Initialize: reset optimizer $\OptMReset{i}$ for all $i \in \Stations$

      \While{optimization is running}
        \State Obtain the current active \gls{STA} combination $\StaSet$
        \OnelineIf{$\StaSet = \emptyset$} Wait $\WaitTime$ and \textbf{continue}
        \State $\AngleSet \gets \OptMSuggest{\StaSet}$
        \State Move antennas to $\AngleSet$ and wait $\WaitTime$
        \OnelineIf{any \gls{STA} in $\StaSet$ is inactive} \textbf{continue} \label{algln:verify_active}
        \ForAll {$i \in \StaSet$}
          \State Calculate average channel capacity $\AvgCapacity_i(\AngleSet)$
        \EndFor

        \If{$\OptMConv$}
          \If{persistent capacity change is detected}
              \State $\OptMReset{i}$ for all $i \in \StaSet$
          \EndIf
        \Else
          \State $\logCap_i \gets \log\AvgCapacity_i(\AngleSet)$ for all $i \in \StaSet$
          \State $\OptMReport{\StaSet}{\AngleSet}{\{\logCap_i\}_{i \in \StaSet}}$
        \EndIf
      \EndWhile
    \end{algorithmic}
  \end{algorithm}
\end{figure}

\begin{figure}[!t]
  \begin{algorithm}[H]
    \caption{Methods of the optimizer $\OptM$} \label{alg:opt_methods}
    \begin{algorithmic}[1]
      \Function{Report $\OptMReportName$}{$\StaSet, \AngleSet, \{\logCap_i\}_{i \in \StaSet}$}
        \ForAll{$i \in \StaSet$}
          \State Update the dataset $\localDS_i \gets \localDS_i \cup \{(\AngleSet, \logCap_i)\}$
          \State Train the surrogate model $\surrogate{i}$ with $\localDS_i$
          \State Increment the counter $\Counter_i$
        \EndFor
        \If{score is improved for $\StaSet$}
          \State $\Fairness_\StaSet^{\mathrm{best}} \gets \frac{1}{\Card{\StaSet}} \sum_{i \in \StaSet} \logCap_i$
          \State $\AngleSetBest_\StaSet \gets \AngleSet$
          \State $\CapBest_{\StaSet,i} \gets \AvgCapacity_i(\AngleSet)$ for all $i \in \StaSet$
        \EndIf
      \EndFunction
    \end{algorithmic}

    \vspace{0.5em}

    \begin{algorithmic}[1]
      \Function{Suggest $\OptMSuggestName$}{$\StaSet$}
        \If{stopping criterion has been reached for $\StaSet$}
          \State \Return $\AngleSetBest_\StaSet$
        \ElsIf{$\exists i \in \StaSet \colon \Counter_i < \InitialPoint$}
            \State \Return an initial orientation for an active link $i$
            \Statex \hspace{2.5em} with $\Counter_i < \InitialPoint$
        \Else
          \State Calculate $\Acq(\AngleSet;\StaSet)$ using \eqref{eq:acquisition}
          \State \Return $\argmax_{\AngleSet}\Acq(\AngleSet;\StaSet)$
        \EndIf
      \EndFunction
    \end{algorithmic}

    \vspace{0.5em}

    \begin{algorithmic}[1]
      \Function{Reset $\OptMResetName$}{$i$}
        \State Reset the dataset $\localDS_{i} \gets \emptyset$,
               the surrogate model $\surrogate{i}$,
        \Statex \hspace{1.2em} and the counter $\Counter_i \gets 0$
        \ForAll{$\StaSet \subseteq \Stations$ with $i \in \StaSet$}
          \State Reset the stopping status for $\StaSet$
          \State Invalidate the best orientation $\AngleSetBest_\StaSet$ and reset $\FairnessBest_\StaSet$
          \State Invalidate $\CapBest_{\StaSet,j}$ for all $j \in \StaSet$
        \EndFor
      \EndFunction
    \end{algorithmic}
  \end{algorithm}
\end{figure}


Algorithm~\ref{alg:proposed} summarizes the extended optimizer operations for time-varying active-\gls{STA} sets.
The methods of $\OptM$ are detailed in Algorithm~\ref{alg:opt_methods}.
Unlike conventional \gls{BO} with a single objective function, the proposed optimizer $\OptM$ receives the current active set $\StaSet$ and the observed log-capacities $\{\logCap_i\}_{i \in \StaSet}$.
The method $\OptMReportName()$ updates the dataset and \gls{GPR} model for each active \gls{STA} $i$ individually, while evaluating the proportional fairness objective for the current active set $\StaSet$.
It also maintains the best objective score $\FairnessBest_\StaSet$ and its corresponding antenna orientation $\AngleSetBest_\StaSet$ for each active set.
Thus, the proposed framework separates link-specific regression models, which are shared across different active sets, from the active-set-specific optimization objectives.

To accommodate the active-set transitions, the other optimizer methods are also extended.
The method $\OptMSuggestName()$ constructs the acquisition function $\Acq(\AngleSet; \StaSet)$ from the regression models corresponding to the current active set $\StaSet$, using \eqref{eq:acquisition}, and selects the next antenna orientation accordingly.
When the active set changes, the acquisition function is reconstructed using the relevant models without discarding previously learned information.
Together, the per-link model updates and active-set-dependent acquisition function enable observation reuse across different active-\gls{STA} combinations without maintaining a separate regression model for each combination.

The proposed method reduces the number of \gls{GPR} models from up to $2^{\Card{\Stations}}-1$ to $\Card{\Stations}$, but it does not remove the need to track the active sets encountered during operation.
However, the main benefit of the proposed approach is to reuse informative capacity observations across active-set transitions and reduce repeated exploration of links already observed.

\section{Evaluation} \label{sec:evaluation}

\subsection{Evaluation Criteria and Baselines}

We evaluated the proposed model-sharing framework for antenna-orientation optimization under changes in the set of active \glspl{STA}.
We first conducted real-value-based simulations to enable reproducible and extensive comparisons of optimization methods and parameter settings, as repeated real-world experiments are time-consuming and subject to channel variability.
Multiple trials under reproducible conditions are also important for fair comparisons, as \gls{BO} may converge to suboptimal solutions.
Based on the simulation results, we selected the parameters of the optimization methods, then conducted the real environment experiments to validate the effectiveness of the proposed method in a real environment.

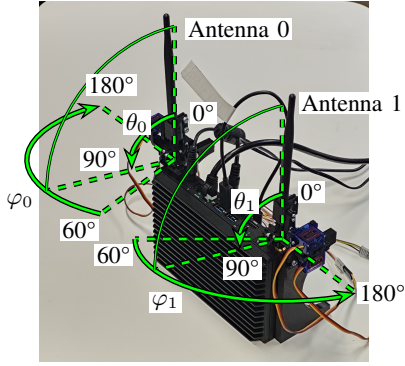
\begin{figure}[!t]
  \centering
  \input{images/angle_def.tex}
  \caption{
    Angle definitions for the azimuth and elevation angles.
  }
  \label{fig:angle_definition}
\end{figure}

We used the device shown in Fig.~\ref{fig:device} both to collect data for the real-value-based simulations and to conduct the real-world experiments.
The device is based on an ASUS NUC 13 Rugged, which can independently control the elevation and azimuth of two antennas using two servo motors for each antenna.
The servo motors are controlled by an Arduino UNO R4 WiFi board.
The elevation and azimuth angles range from 0 to 90 degrees and from 60 to 180 degrees, respectively, as shown in Fig.~\ref{fig:angle_definition}.
PicoScenes \cite{jiang2022eliminating} was installed on the proposed receiver and the transmitting devices to acquire \gls{RSSI} and \gls{CSI} measurements.
We utilized two devices as \TXpl: one is an ASUS NUC 13 Rugged (\TXa) and the other is an Intel NUC 12 Pro Kit (\TXb).
During the evaluation, these \TXpl periodically inject \gls{CSI} measurement packets to the proposed device, which acts as a receiver.
The center frequency and bandwidth of the \WiFi channel are set to 5.57\,GHz and 160\,MHz, respectively.

We compared the following five methods: (i) fixed antenna orientation ({\small\textsf{Fixed}}), (ii) random search ({\small\textsf{Random}}), (iii) state-specific \gls{BO} \cite{taya2026globecom} ({\small\textsf{SS BO}}), and (iv) model-sharing framework with \gls{cUCB} \cite{jain2023bayesian} ({\small\textsf{MS (\gls{cUCB})}}), and (v) model-sharing framework with \gls{propUCB} defined in \eqref{eq:acquisition} ({\small\textsf{MS (\gls{propUCB})}}).
Note that the optimizer optimizes the proportional fairness defined in \eqref{eq:fairness} but this section presents the geometric mean of the channel capacity \eqref{eq:geomean} as the evaluation metric because it is a monotonic transformation of the objective and is easier to interpret.
When comparing the convergence speed of the methods, we used normalized optimality, which is defined as the ratio of the geometric mean of the channel capacity to the per-state optimum in real-value-based simulation.

\subsection{Real-Value-Based Simulation}
To compare the performance of the proposed methods, we first conduct a real-value-based simulation.
Although one run of antenna-orientation optimization takes only a few minutes, conducting sufficiently many trials in a real environment is time-consuming and difficult to reproduce.
Therefore, we first measured the channel capacities in grid points of the antenna orientations in a real environment, and then conducted a simulation using the measured channel capacities.

\begin{figure}[!t]
  \centering
  \subfloat[\roomA: small meeting room]{
    \resizebox{0.19\textwidth}{!}{\input{images/location_small.tex}}
    \label{fig:location_roomA}
  }
  \hspace{0.2em}
  \subfloat[\roomB: class room]{
    \resizebox{0.26\textwidth}{!}{\input{images/location.tex}}
    \label{fig:location_roomB}
  }\\
  \subfloat[photo of \roomB]{
    \includegraphics[width=0.35\textwidth]{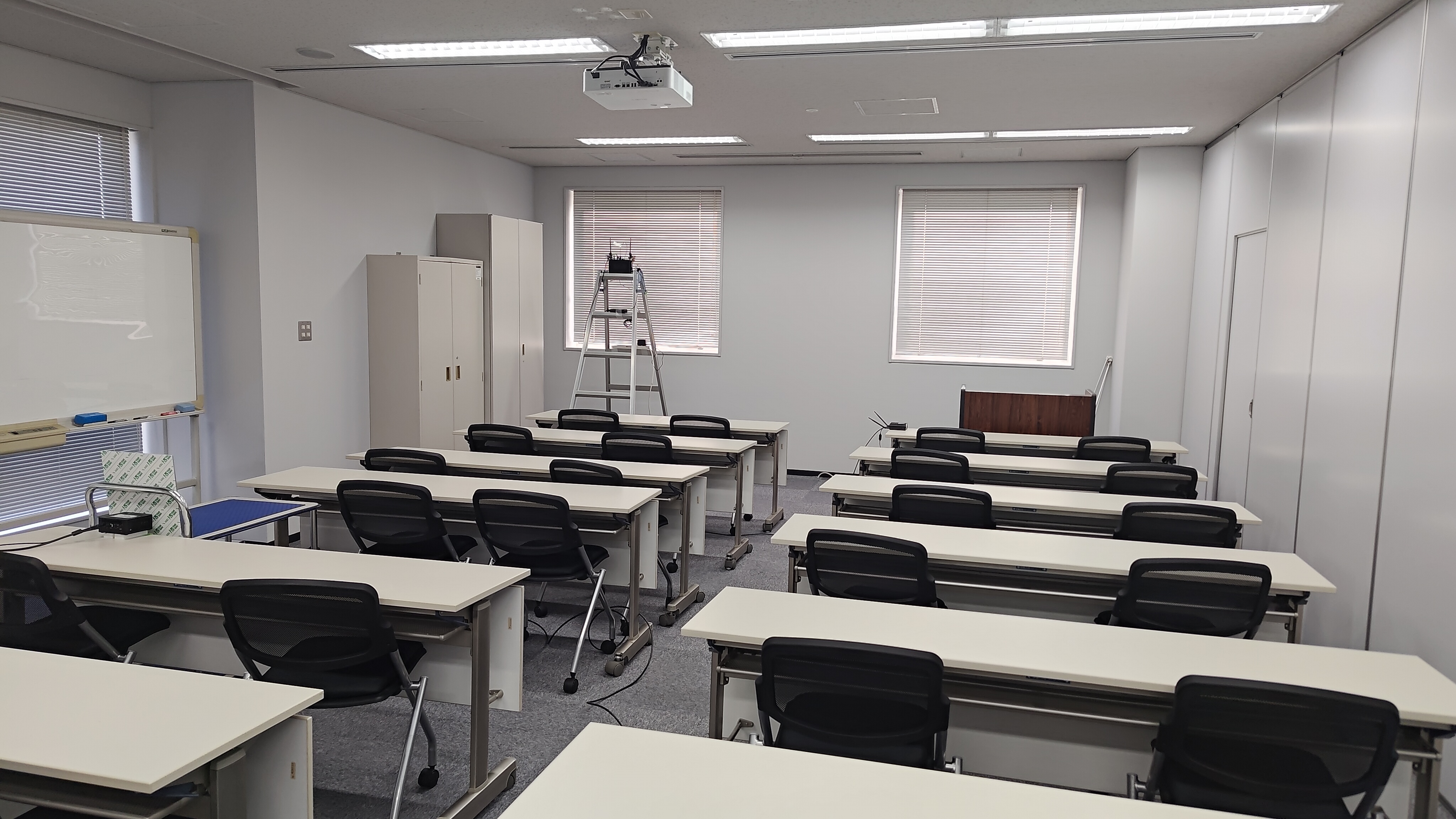}
    \label{fig:location_roomB_photo}
  }
  \caption{
    Layouts of the measurement environments.
    The proposed device (\RX) and two \TXpl (\TXa and \TXb) were mounted on a ladder at a height of 1.7\,m and on desks at a height of 0.7\,m, respectively.
    In \roomB, the two \TXpl are placed at different locations in two settings: \SetBlock and \SetLos.
  }
  \label{fig:location}
\end{figure}

The values of the channel capacity $\AvgCapacity_i(\AngleSet)$ used in the simulation are obtained in the following manner.
We evaluated the proposed device in two different rooms, \roomA and \roomB, as shown in Fig.~\ref{fig:location}.
The proposed device (\RX) was mounted on a ladder at a height of 1.7\,m, while the two \TXpl (\TXa and \TXb) were placed on desks at a height of 0.7\,m.
In \roomB, we measured two settings \SetBlock and \SetLos, where the two \glspl{STA} are placed at different locations.
In the setting \SetBlock, an aluminum plate with 300\,mm\,$\times$\,300\,mm\,$\times$\,5\,mm is placed in front of the \TXb to block the direct path.

During data collection, the two \TXpl periodically transmitted measurement frames to the receiver; interference from other devices was assumed negligible.
The channel capacity $\AvgCapacity_i(\AngleSet)$ is measured for each \gls{STA} $i$ and each antenna orientation $\AngleSet$ in the two layouts.
$\AngleSet$ is discretized in 15-degree increments for both azimuth and elevation angles, each ranging from 60 to 180 and 0 to 90 degrees, respectively, resulting in a total of 3,969 configurations for two antennas.
At each point, we measured at least three valid \gls{CSI} and \gls{RSSI} values for two \TXpl, and then calculated the average channel capacity $\AvgCapacity_i(\AngleSet)$.
When calculating channel capacities, we set temperature $\Temperature=300$\,K and a receiver noise figure of 7\,dB, which are typical values for \WiFi receivers \cite{merlin2015tgax}.
The noise density is calculated as $\NoiseDensity=\Boltzman \Temperature \NoiseFactor$, where $\Boltzman$ and $\NoiseFactor$ denote the Boltzmann constant and the linear noise factor, respectively.

\begin{figure}[!t]
  \centering
  \includegraphics[width=0.40\textwidth]{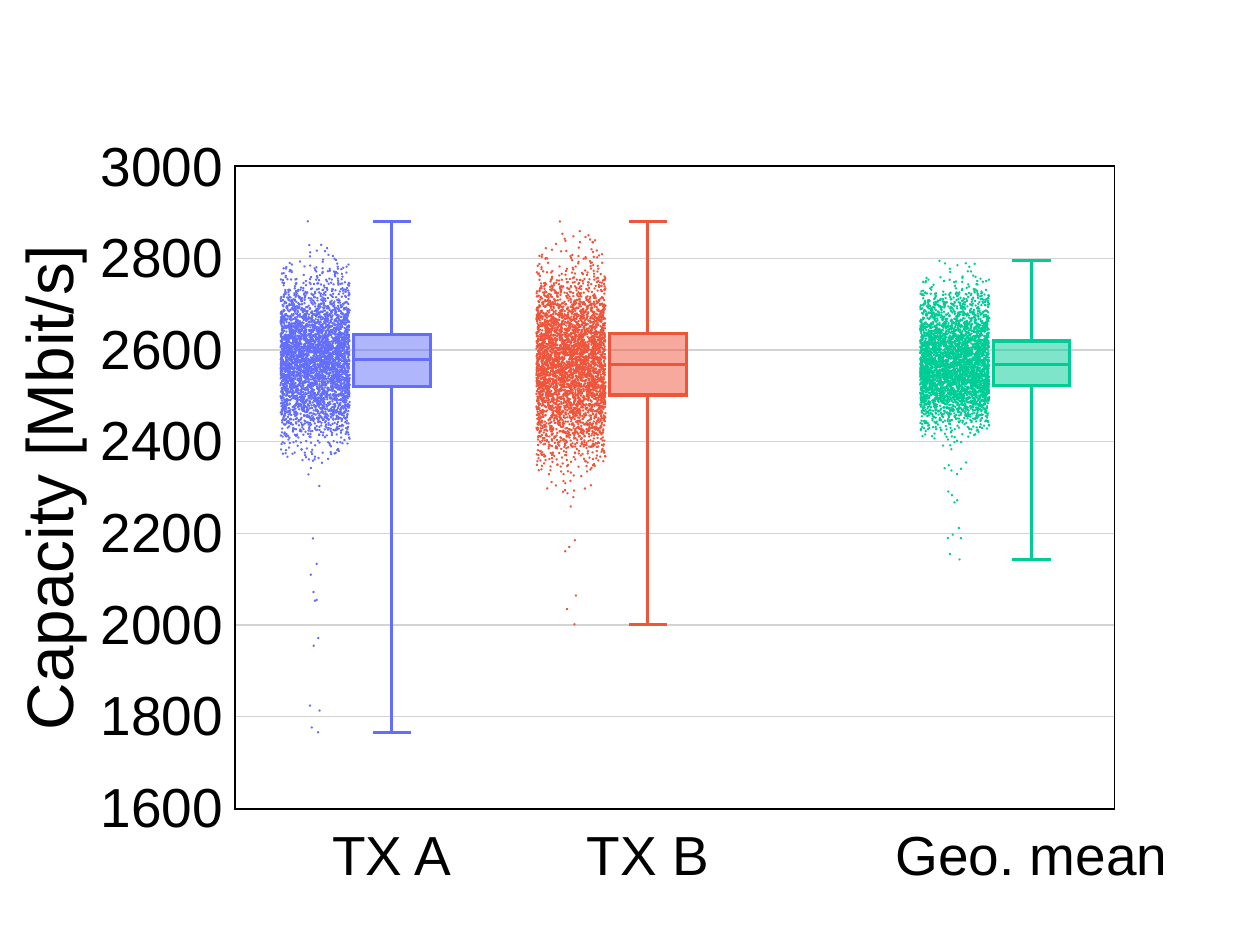}
  \caption{
  Distribution of measured channel capacities over all antenna-orientation configurations in the grid for \TXa, \TXb, and their geometric mean.
  The boxes represent the IQR, with the median indicated by a horizontal line.
  The whiskers extend to the minimum and maximum observed values, and individual points represent all measured configurations without outlier exclusion.
  }
  \label{fig:real_value_sim_small_stats}
\end{figure}

\begin{figure}[!t]
  \centering
  \subfloat[\TXa $\rightarrow$ Antenna 0]{
    \includegraphics[width=\subfigWidth]{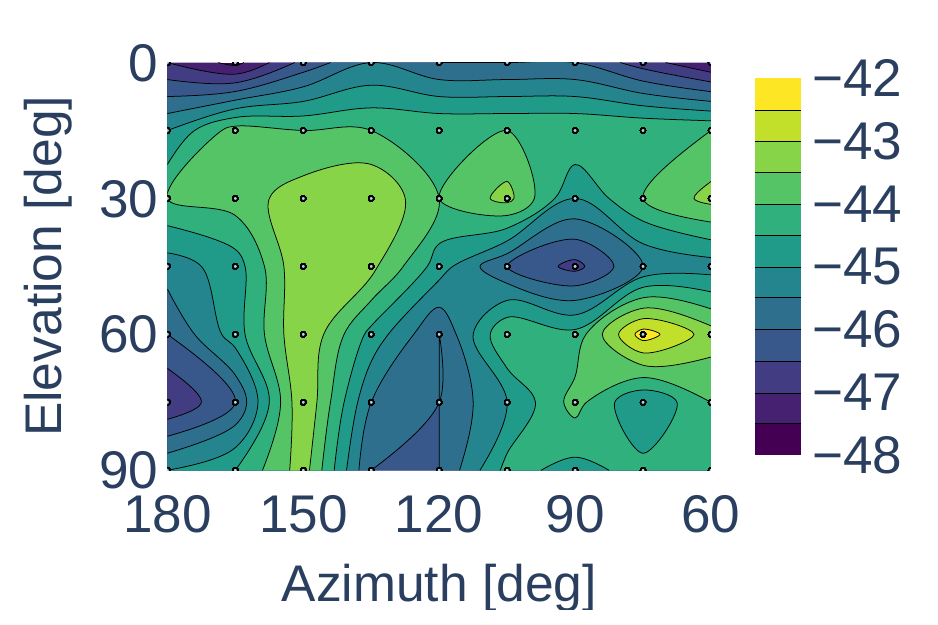}
    \label{fig:real_value_sim_contour_TXa0}
  }
  \subfloat[\TXa $\rightarrow$ Antenna 1]{
    \includegraphics[width=\subfigWidth]{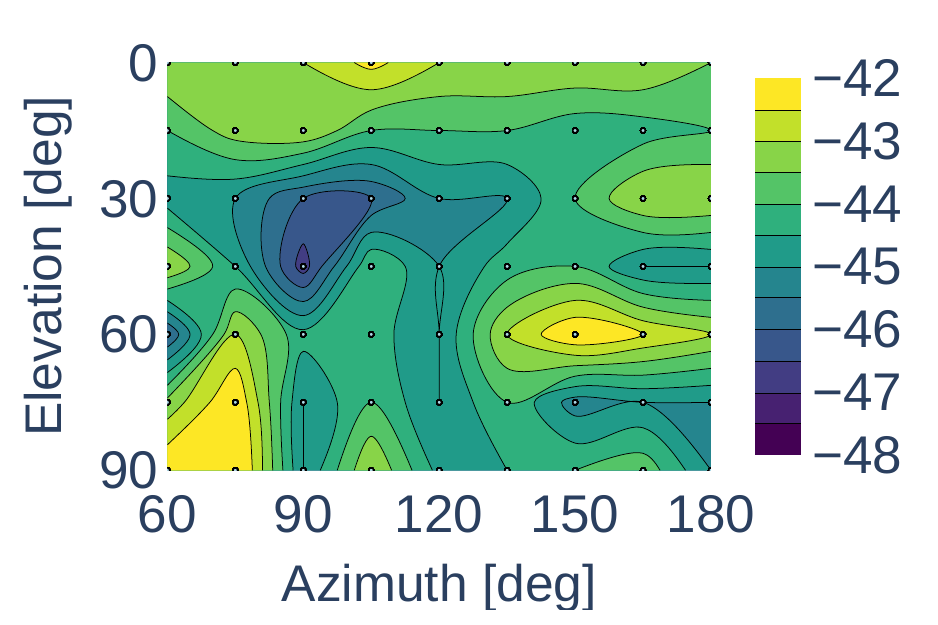}
    \label{fig:real_value_sim_contour_TXa1}
  }\\
  \vspace{-1em}
  \subfloat[\TXb $\rightarrow$ Antenna 0]{
    \includegraphics[width=\subfigWidth]{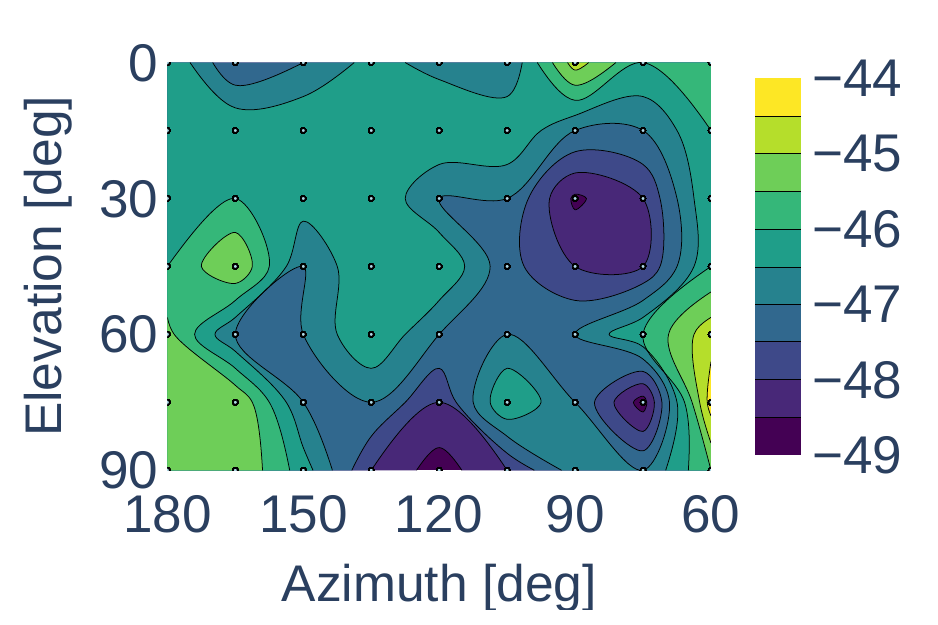}
    \label{fig:real_value_sim_contour_TXb0}
  }
  \subfloat[\TXb $\rightarrow$ Antenna 1]{
    \includegraphics[width=\subfigWidth]{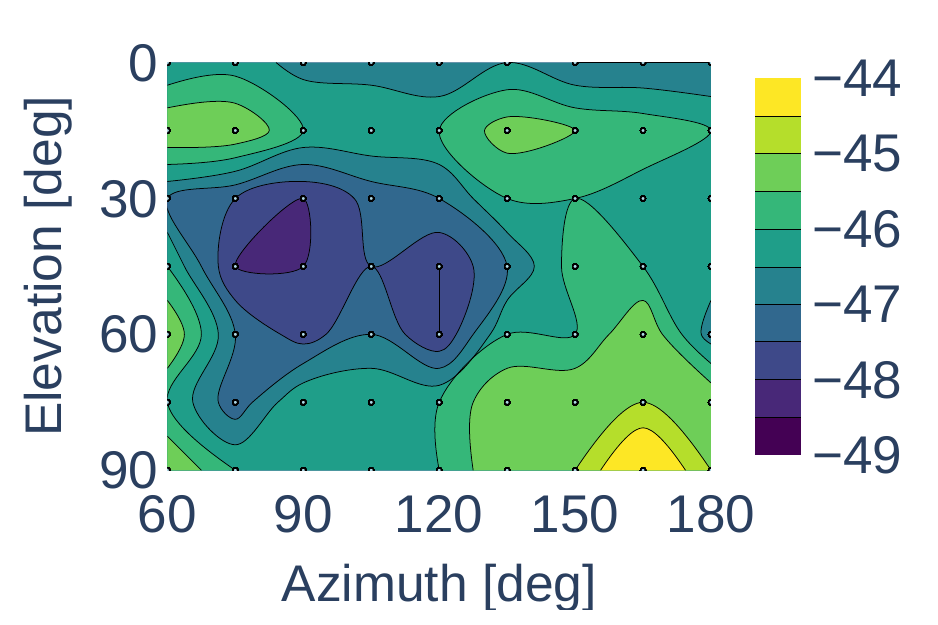}
    \label{fig:real_value_sim_contour_TXb1}
  }\\
  \vspace{-1em}
  \subfloat[\TXc $\rightarrow$ Antenna 0]{
    \includegraphics[width=\subfigWidth]{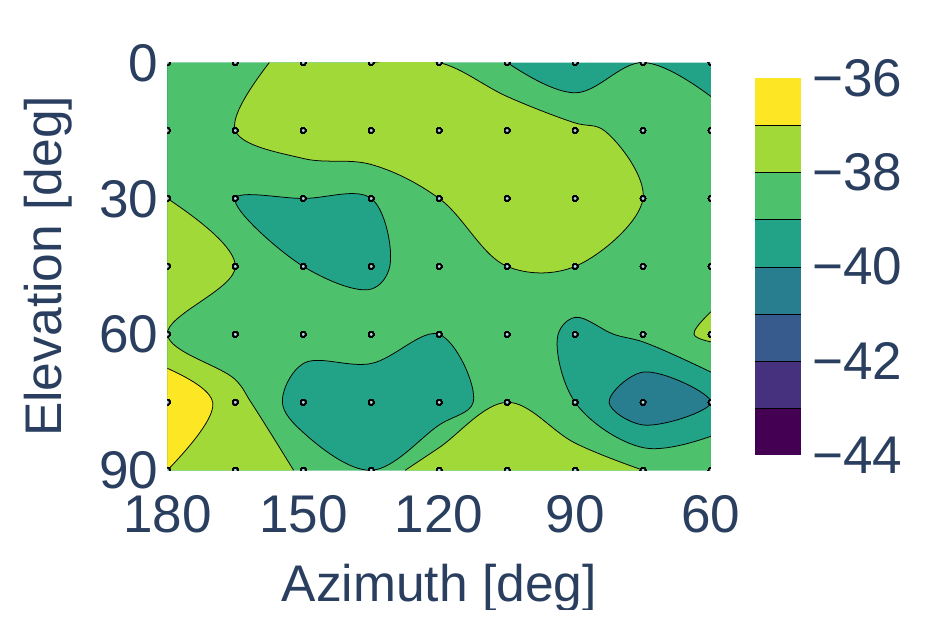}
    \label{fig:real_value_sim_contour_TXc0}
  }
  \subfloat[\TXc $\rightarrow$ Antenna 1]{
    \includegraphics[width=\subfigWidth]{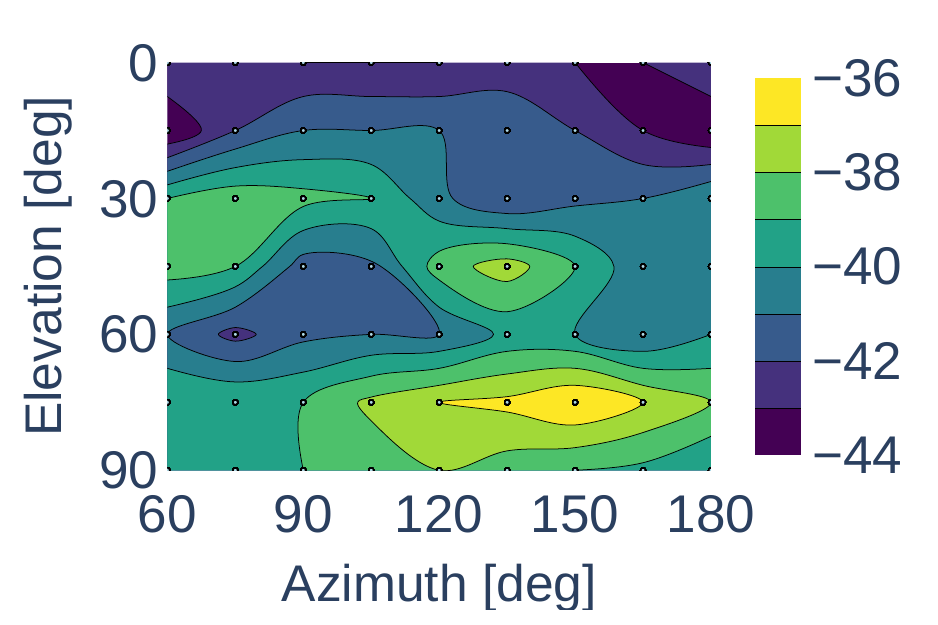}
    \label{fig:real_value_sim_contour_TXc1}
  }\\
  \vspace{-1em}
  \subfloat[\TXd $\rightarrow$ Antenna 0]{
    \includegraphics[width=\subfigWidth]{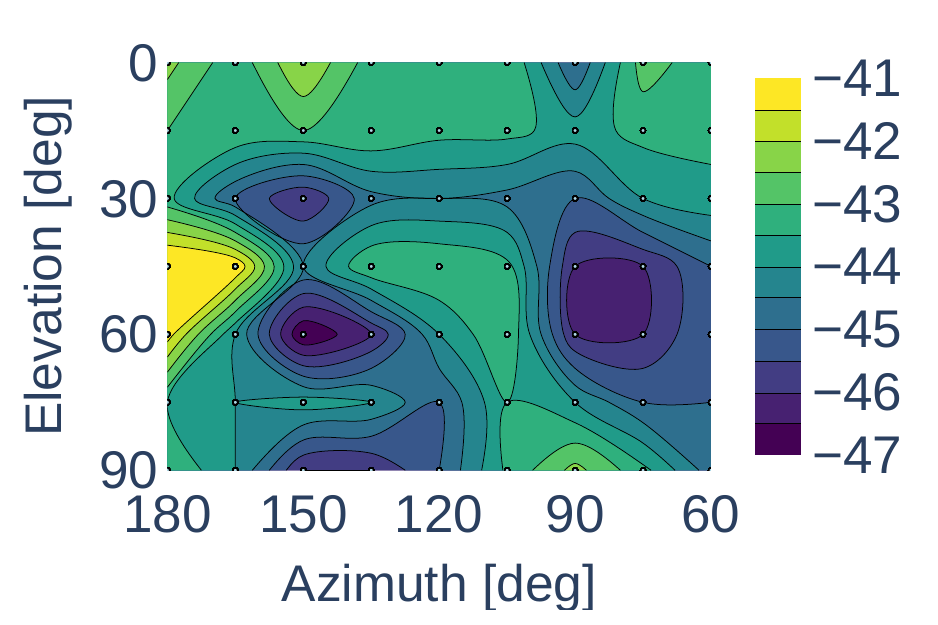}
    \label{fig:real_value_sim_contour_TXd0}
  }
  \subfloat[\TXd $\rightarrow$ Antenna 1]{
    \includegraphics[width=\subfigWidth]{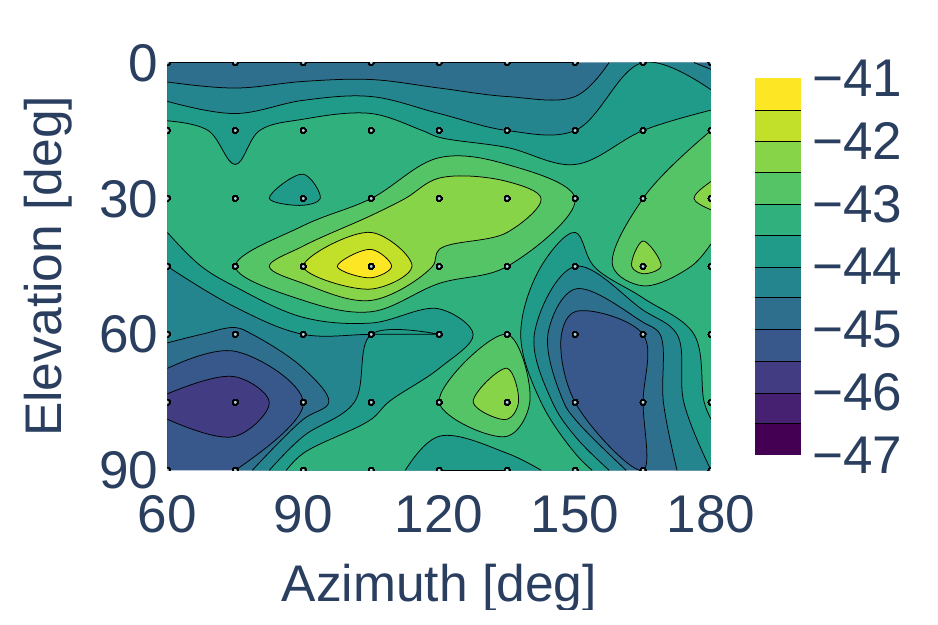}
    \label{fig:real_value_sim_contour_TXd1}
  }
  \caption{
    Contour plots of the measured \gls{RSSI} values from each \gls{STA} to each antenna.
    Dots indicate the measured grid points.
    Different \glspl{STA} exhibit distinct spatial patterns, and the optimal antenna orientations for each \gls{STA} are not necessarily the same.
  }
  \label{fig:real_value_sim_contour}
\end{figure}

After the measurement, we also constructed a virtual four-link dataset by combining these \roomBblock and \SetLos.
Note that this merged dataset is not a real measurement of four links simultaneously, thus it does not reflect the interference between the two \TXpl and propagation changes due to the presence of multiple \glspl{STA} themselves.
However, this dataset is useful for evaluating the optimization methods when the number of \glspl{STA} increases.
In the merged four-link dataset, \TXa and \TXb measured in \roomBlos are renamed as \TXc and \TXd, respectively, while those measured in \roomBblock are treated as they are.

In these simulations, we formulate a discrete optimization problem that maximizes the geometric-mean capacity $\GeoMCapacity_{\StaSet}(\AngleSet)$ of the active \glspl{STA} over the measured orientations.
Although the optimal orientation can be found by exhaustive search in this discrete optimization problem, we limit the number of searches for each method to compare the optimization efficiency.
Note that in the real-value-based simulation, the maximization of the acquisition function is performed by evaluating the acquisition function at all grid points of the measured antenna orientations, rather than using gradient-based optimization methods, which is a different setting from the following real experiments.

Fig.~\ref{fig:real_value_sim_small_stats} shows the distributions of the measured channel capacities over all antenna-orientation configurations in the grid.
Although the \gls{IQR} is only 97.6\,Mbit/s, indicating that the middle 50\% of configurations exhibit a relatively narrow capacity range, the geometric mean of the channel capacities ranges from 2,171 to 2,795\,Mbit/s, corresponding to a 28.7\% increase from the worst to the best configuration.
These results demonstrate that antenna orientation can substantially affect the channel capacity, motivating the optimization of antenna orientations to identify favorable configurations for channel quality improvements and energy efficiency.

The contour plots of the \gls{RSSI} from each \TXpl to each antenna are shown in Fig.~\ref{fig:real_value_sim_contour}.
It can be observed that the \gls{RSSI} values vary across different \TXpl even for the same orientation, indicating that the optimization of antenna orientations for multiple links is a nontrivial task.
In addition, the two antennas exhibit different spatial patterns.
Their separation, on the order of half a wavelength, can produce different multipath fading responses; this complexity motivates our use of black-box optimization rather than an explicit geometric propagation model.

\begin{figure}[!t]
  \centering
  \subfloat[Comparison with $\beta$ for \gls{cUCB}]{
    \includegraphics[width=\subfigWidth]{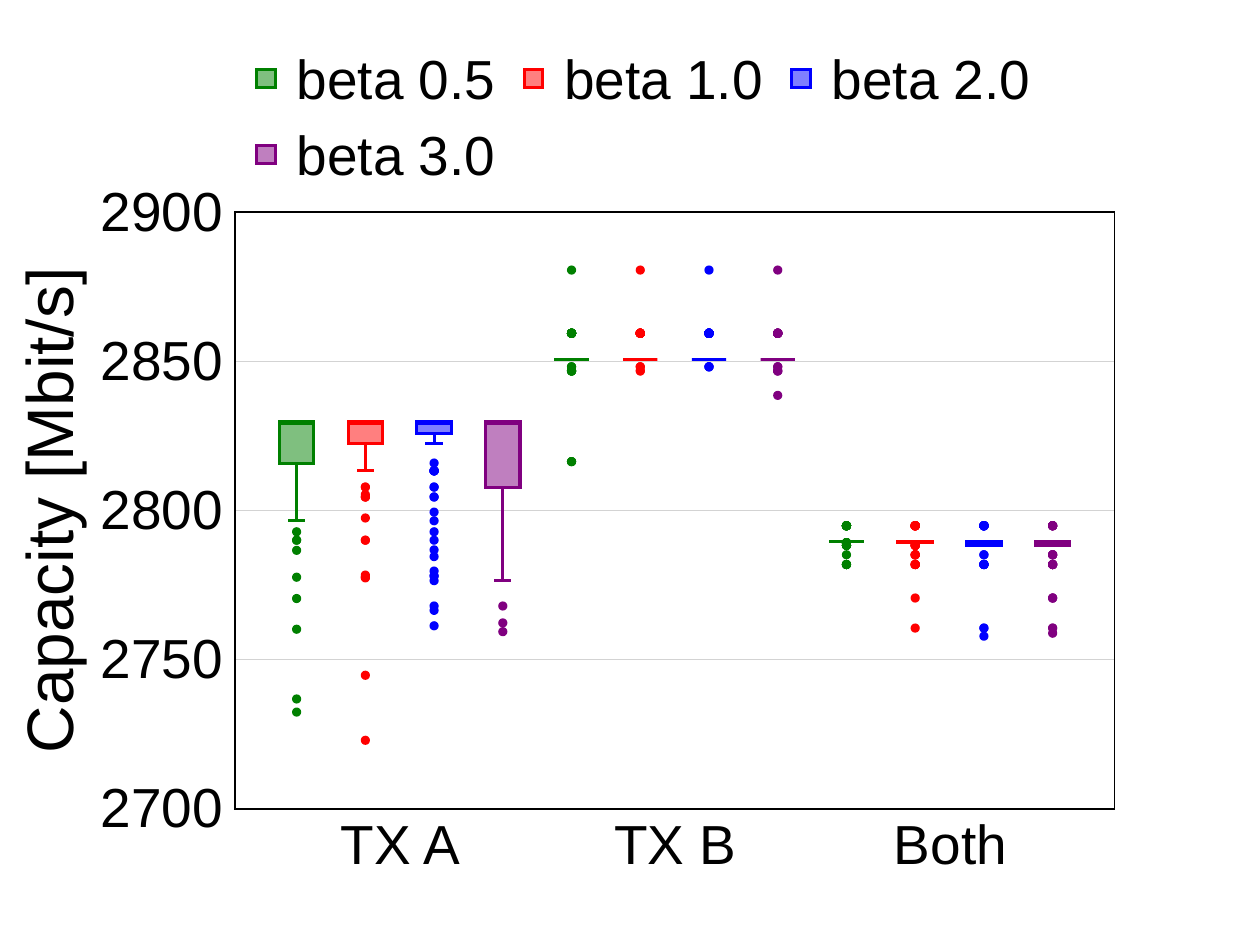}
    \label{fig:real_value_sim_ablation_beta}
  }
  \subfloat[Comparison with $\InitialPoint$ for \gls{propUCB}]{
    \includegraphics[width=\subfigWidth]{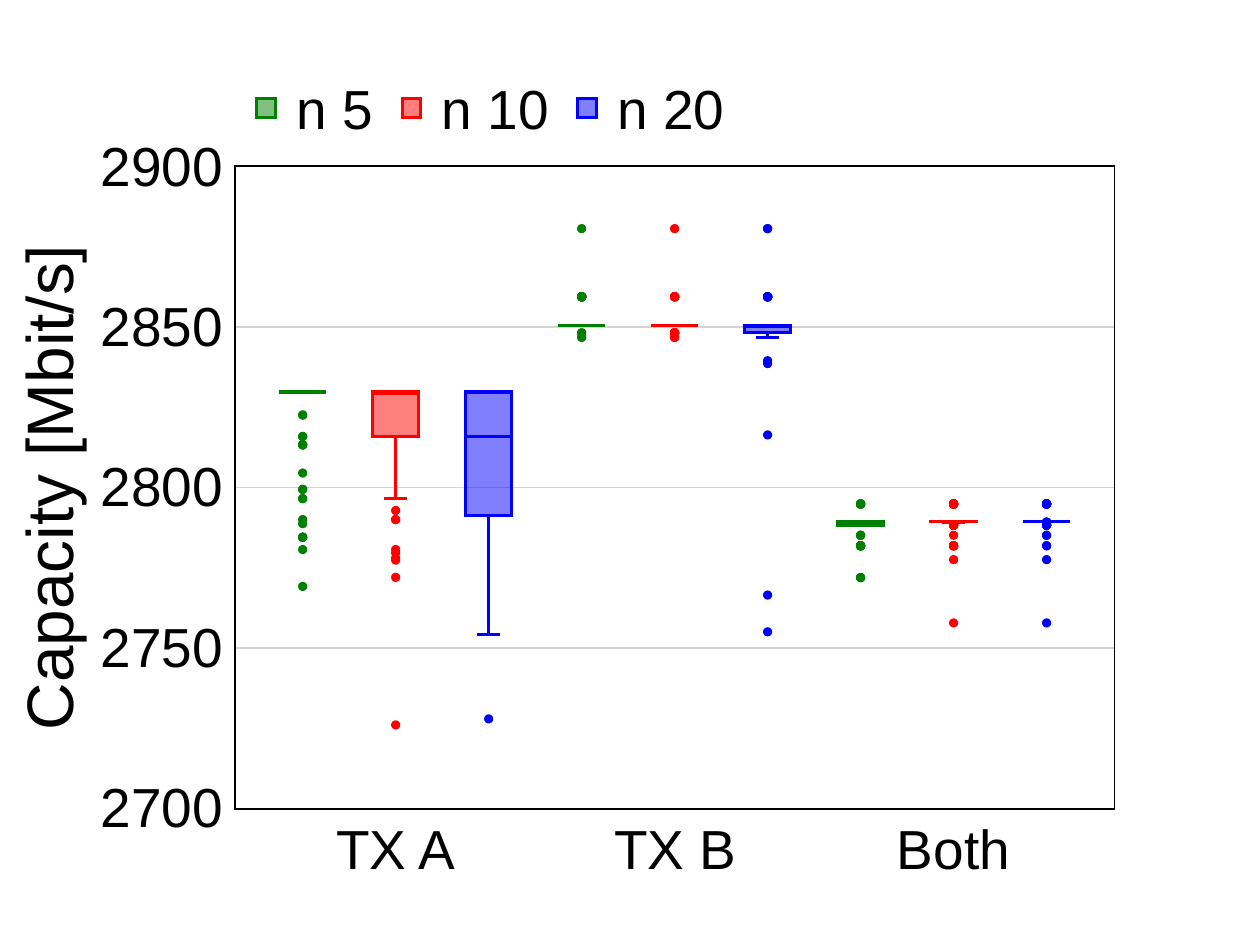}
    \label{fig:real_value_sim_ablation_initn}
  }\\
  \subfloat[Comparison with $\ucbCoef$ for \gls{propUCB}]{
    \includegraphics[width=\subfigWidth]{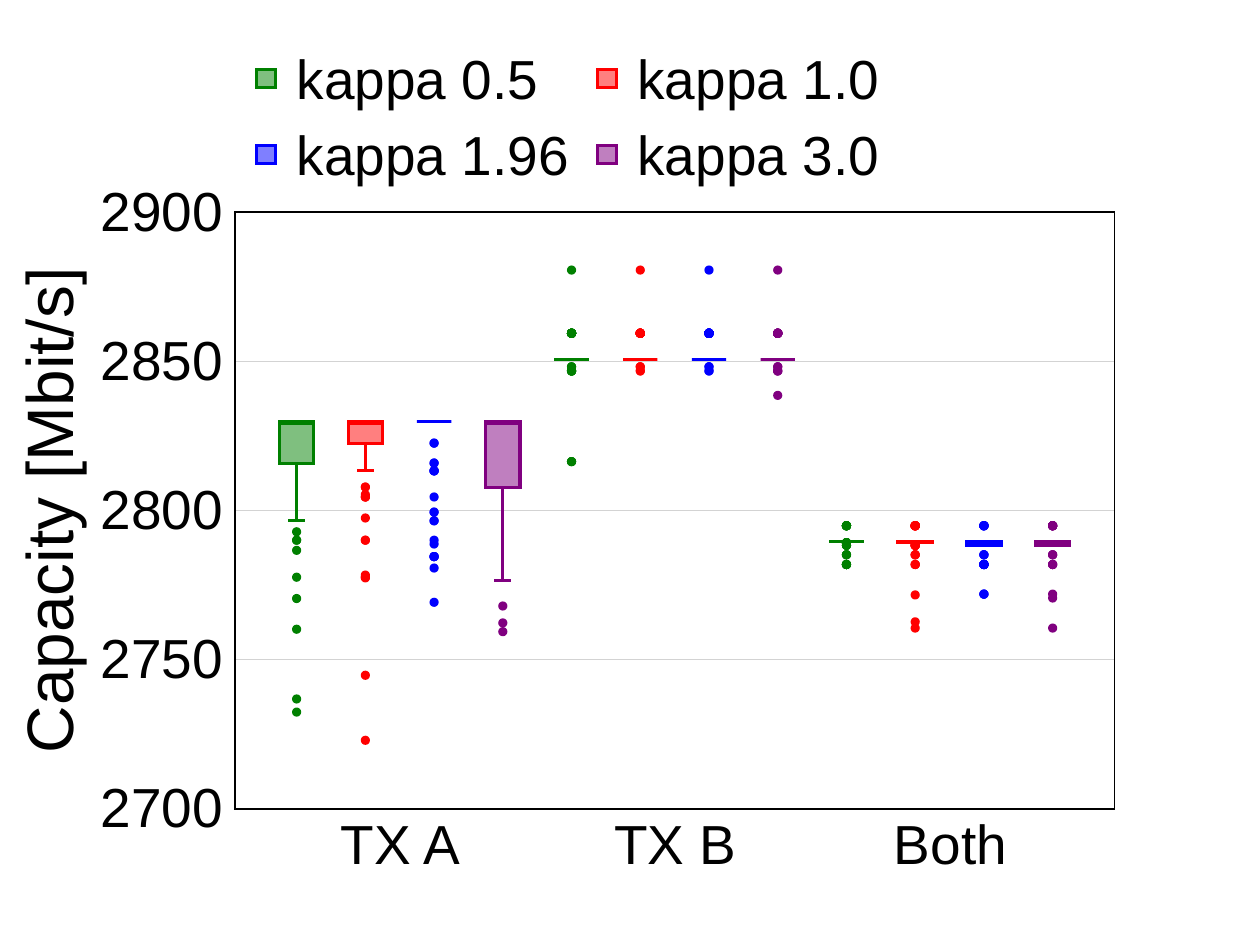}
    \label{fig:real_value_sim_ablation_kappa}
  }
  \caption{
    Comparison of parameters for \gls{cUCB} and \gls{propUCB}.
    The boxes represent the IQR, with the median indicated by a line inside each box.
    The whiskers follow Tukey’s 1.5$\times$IQR rule, and outliers are plotted as individual points.
  }
  \label{fig:real_value_sim_ablation}
\end{figure}

Fig.~\ref{fig:real_value_sim_ablation} shows the comparison among hyper parameters.
In this evaluation, only \TXa was active for the first 50 iterations, followed by \TXb alone for the next 50 iterations.
After 50 evaluations with only \TXb active, both \TXpl became active, and their geometric-mean capacity was evaluated for another 50 iterations.
We denote this sequence \seqA.
The parameters not compared are fixed as $\InitialPoint=5$ and $\ucbCoef=1.96$.
Although $\beta=2.0$ shows stable performance for \TXa in Fig.~\ref{fig:real_value_sim_ablation_beta}, the median values are comparable.
Therefore, we set $\beta=1.0$, which is the default value in \cite{jain2023bayesian} for the following evaluations.
For \gls{propUCB}, Figs.~\ref{fig:real_value_sim_ablation_initn}~and~\ref{fig:real_value_sim_ablation_kappa} demonstrate small deviation with initial points $\InitialPoint=5$ and $\ucbCoef=1.96$ especially for \TXa; thus we used these values for the subsequent evaluations.

\begin{figure}[!t]
  \centering
  \subfloat[\roomA\\\seqA]{
    \includegraphics[width=\subfigWidth]{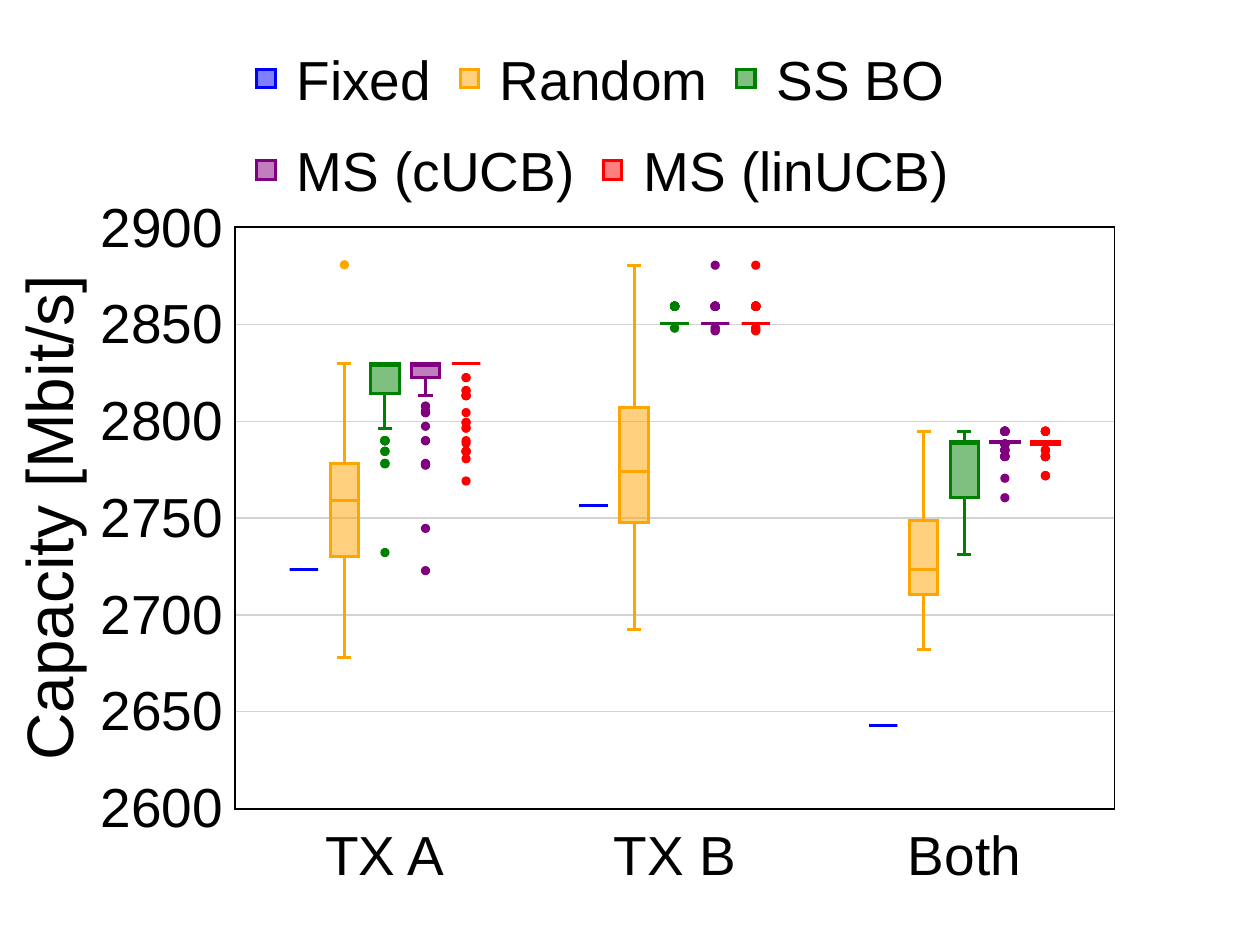}
    \label{fig:real_value_sim_small_all50}
  }
  \subfloat[\roomA\\\seqB]{
    \includegraphics[width=\subfigWidth]{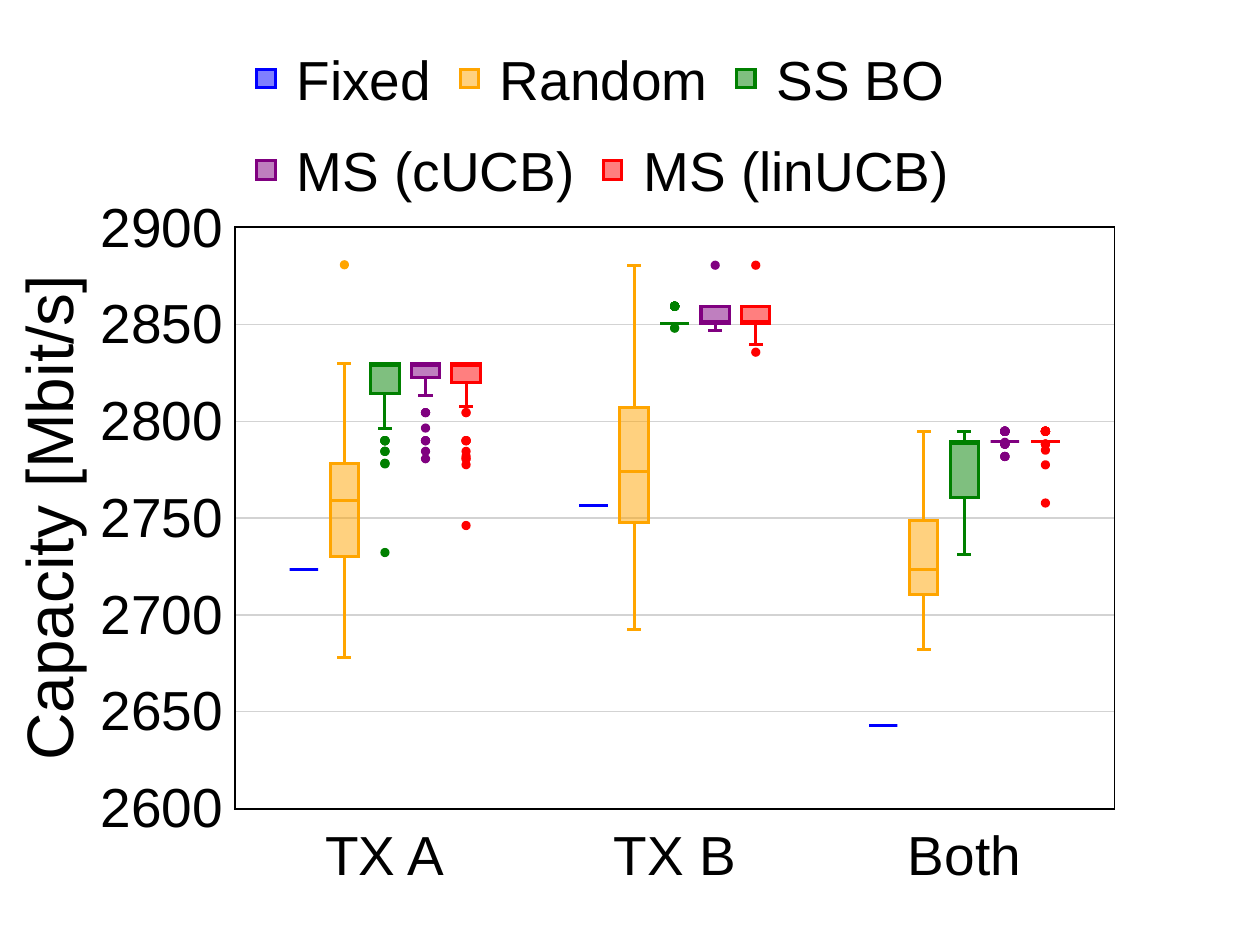}
    \label{fig:real_value_sim_small_10x5}
  }\\

  \subfloat[\roomBblock\\\seqA]{
    \includegraphics[width=\subfigWidth]{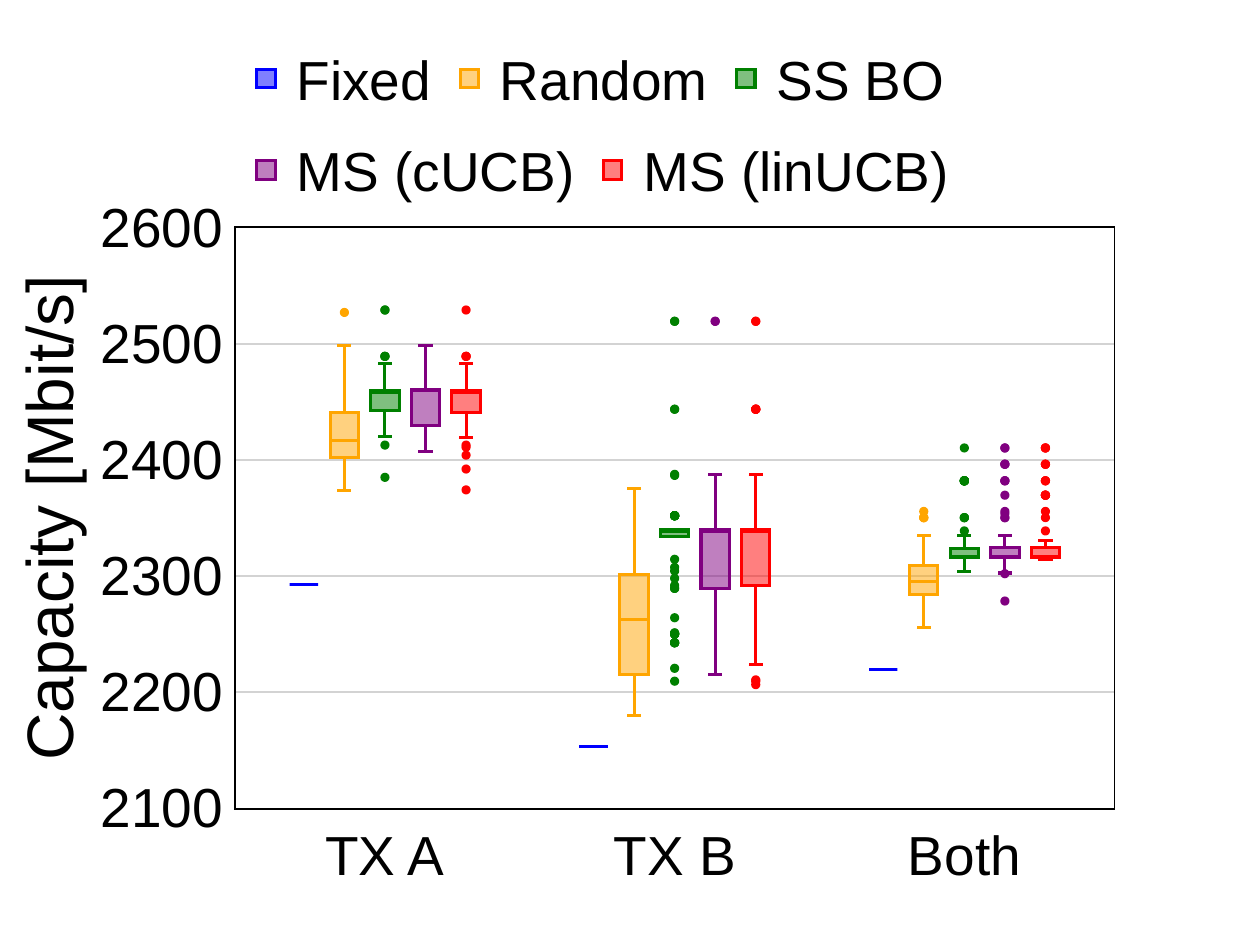}
    \label{fig:real_value_sim_block_all50}
  }
  \subfloat[\roomBblock\\\seqB]{
    \includegraphics[width=\subfigWidth]{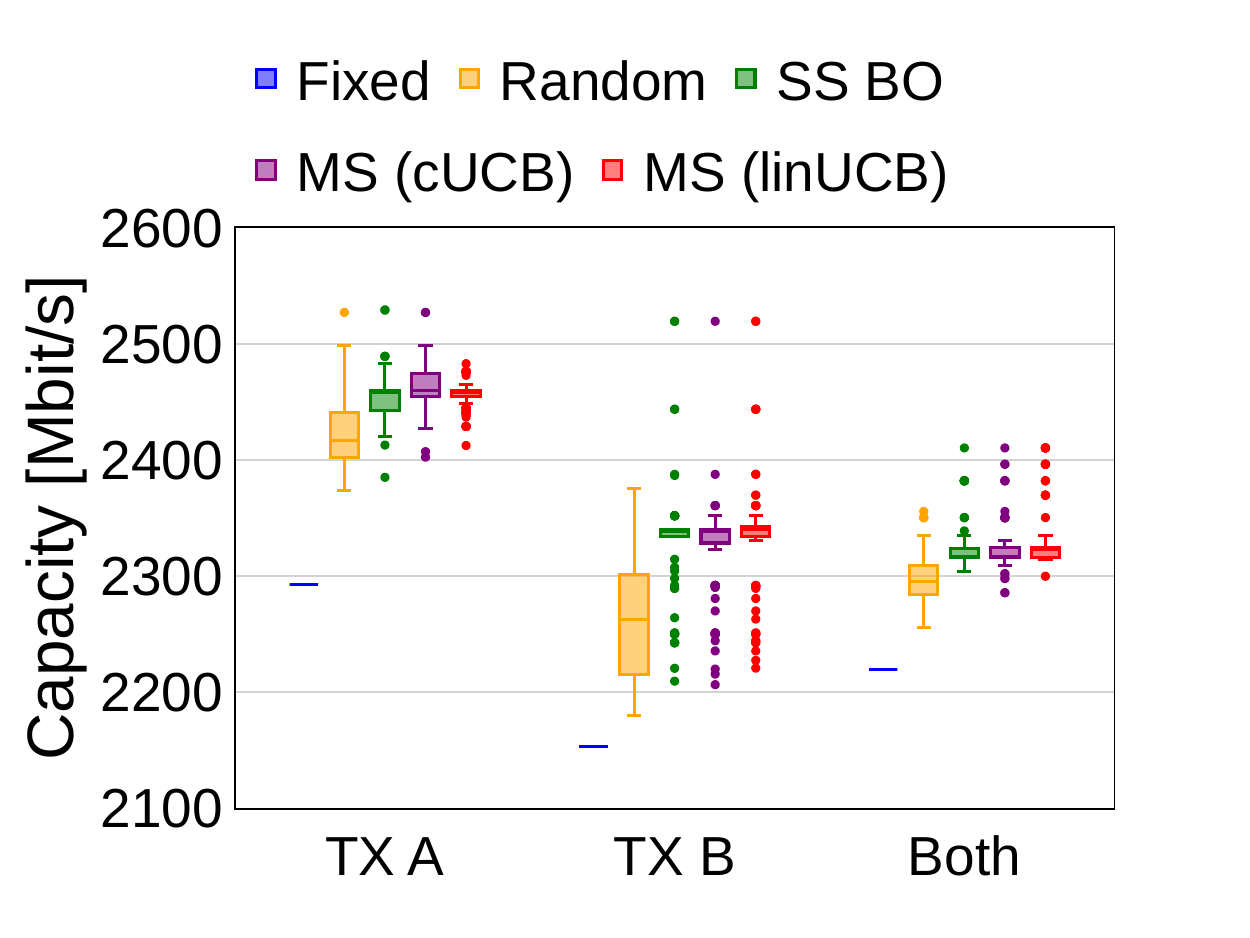}
    \label{fig:real_value_sim_block_10x5}
  }\\

  \subfloat[\roomBlos\\\seqA]{
    \includegraphics[width=\subfigWidth]{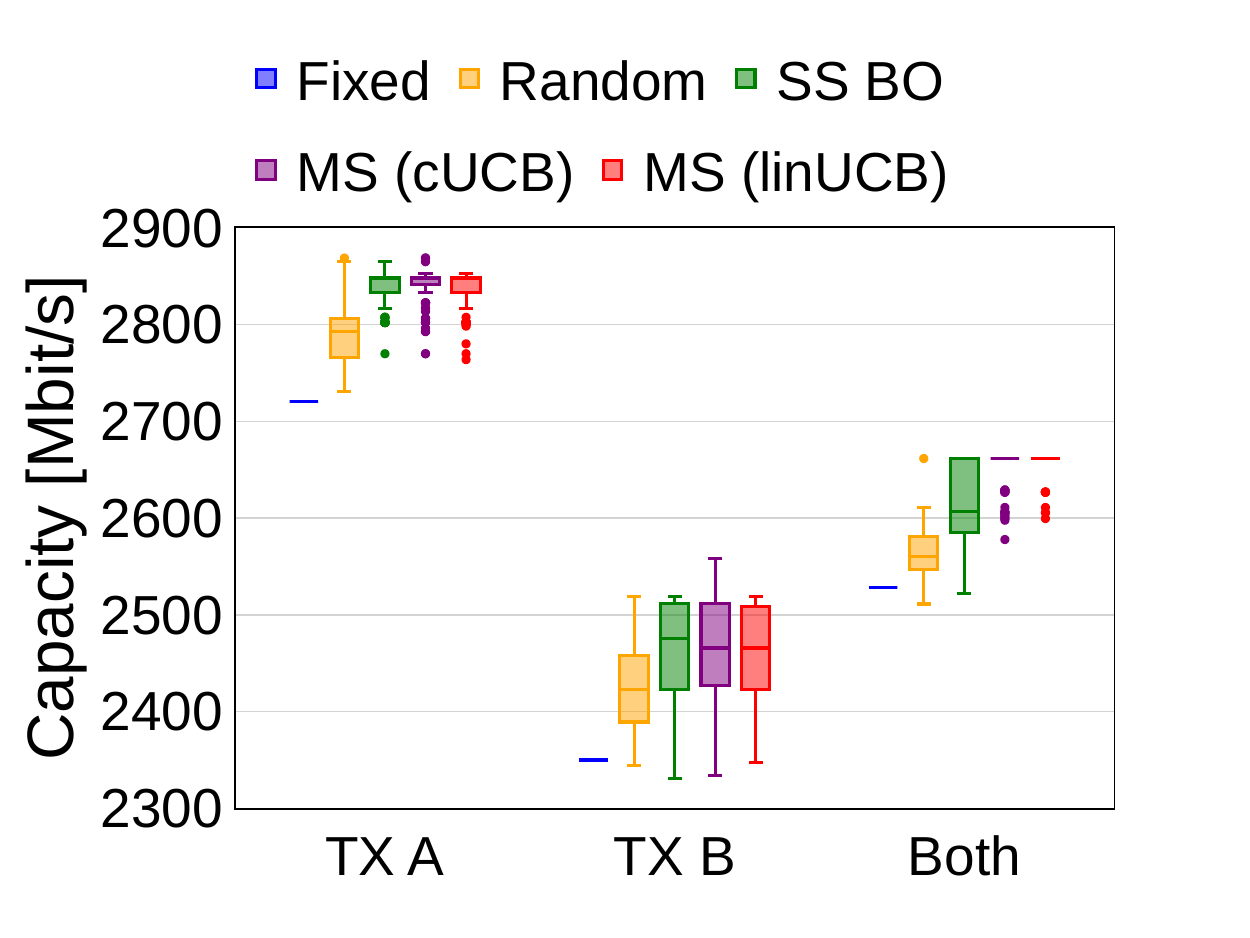}
    \label{fig:real_value_sim_los_all50}
  }
  \subfloat[\roomBlos\\\seqB]{
    \includegraphics[width=\subfigWidth]{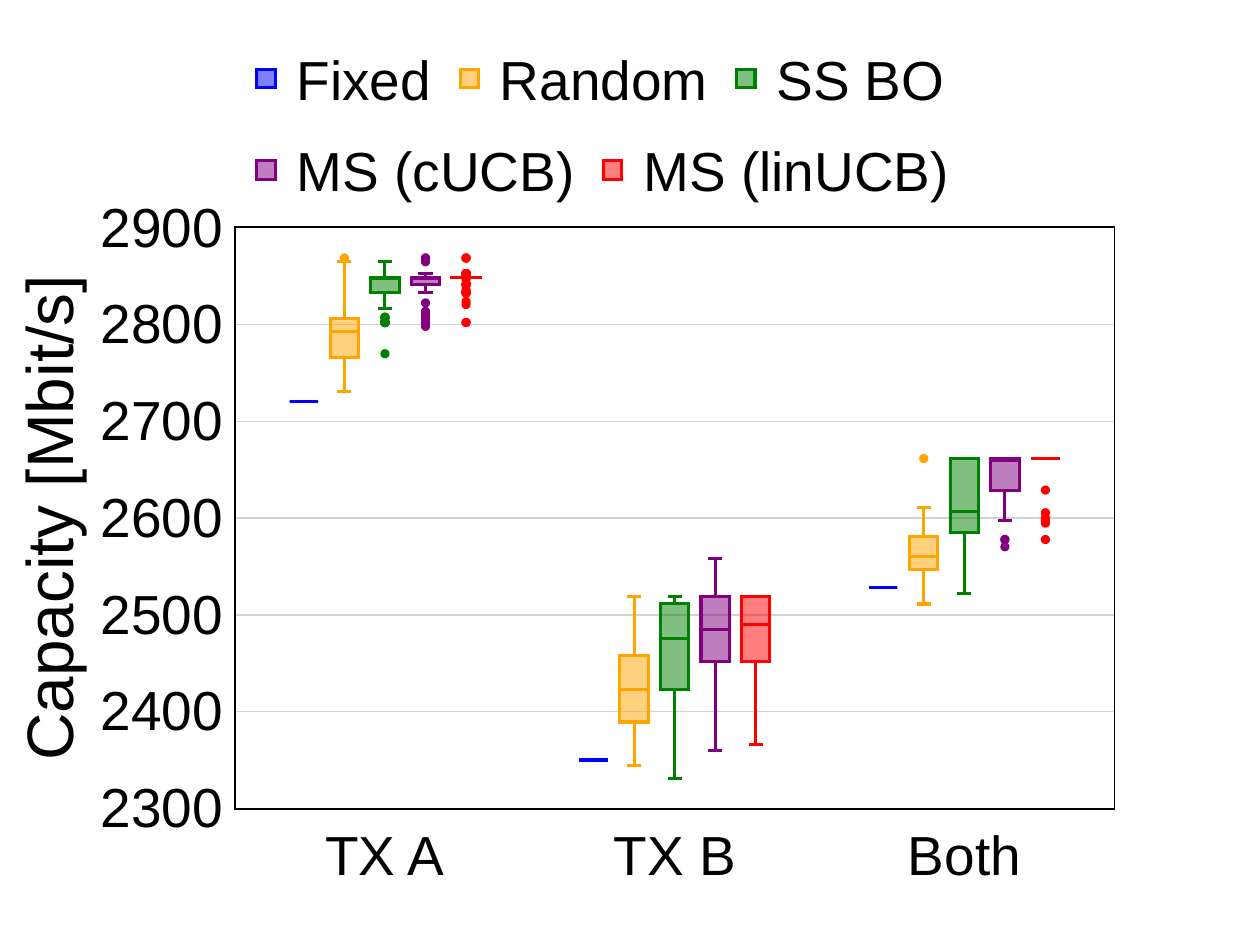}
    \label{fig:real_value_sim_los_10x5}
  }

  \caption{
    Comparison of optimization methods.
    The boxes, lines inside the boxes, and individual points represent the IQR, median values, and outliers.
    The proposed model-sharing framework achieved larger capacities in all layouts.
    Two acquisition functions \gls{cUCB} and \gls{propUCB} show comparable performances.
  }
  \label{fig:real_value_sim_small}
\end{figure}

Fig.~\ref{fig:real_value_sim_small} compares the optimization results for the geometric mean of the channel capacity with various methods.
For the fixed-orientation baseline, we show the best values among four antenna orientations $\AngleSet=(\theta_0,\varphi_0,\theta_1,\varphi_1)$, where $\theta$ represents the elevation angle and $\varphi$ represents the azimuth angle.
These orientations are $(0^\circ,90^\circ,0^\circ,90^\circ)$, $(90^\circ,90^\circ,90^\circ,90^\circ)$, $(90^\circ,180^\circ,90^\circ,180^\circ)$, and $(0^\circ,180^\circ,0^\circ,180^\circ)$.
We evaluated two sequences: \seqA\ and \seqB; in the latter, each of the three active sets is evaluated for 10 iterations, and the cycle is repeated five times.
Although state-specific \gls{BO} outperformed random search, its results for the both-TX-active state exhibited a larger \gls{IQR} than those of the model-sharing methods in \roomA and \roomBlos.
In contrast, the model-sharing methods more consistently identify high-capacity antenna orientations.
This improvement is attributed to the reuse of link observations collected during both single-TX and dual-TX active states.
In \roomBblock, where the direct path to \TXb is blocked by the aluminum plate, the optimized capacities for \TXb exhibit a wider distribution, with a median value substantially lower than the best observed capacity.
This suggests that the antenna orientation optimization is more difficult for blocked path than \gls{LOS} path.

\begin{figure}[!t]
  \centering
  \subfloat[Measured capacity distributions]{
    \includegraphics[width=0.30\textwidth]{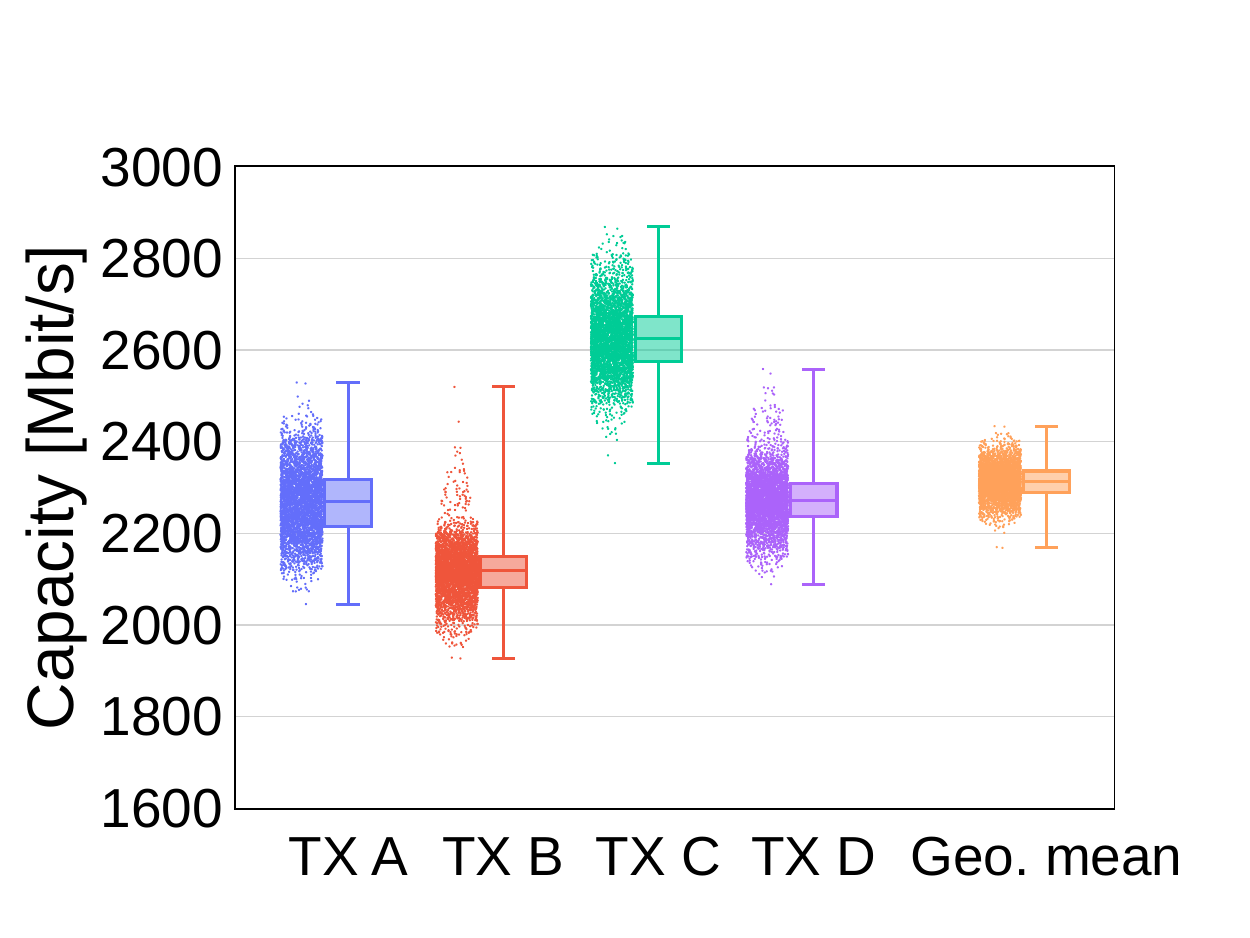}
    \label{fig:real_value_sim_large_merge_stats}
  }\\
  \subfloat[\seqMA]{
    \includegraphics[width=0.48\textwidth]{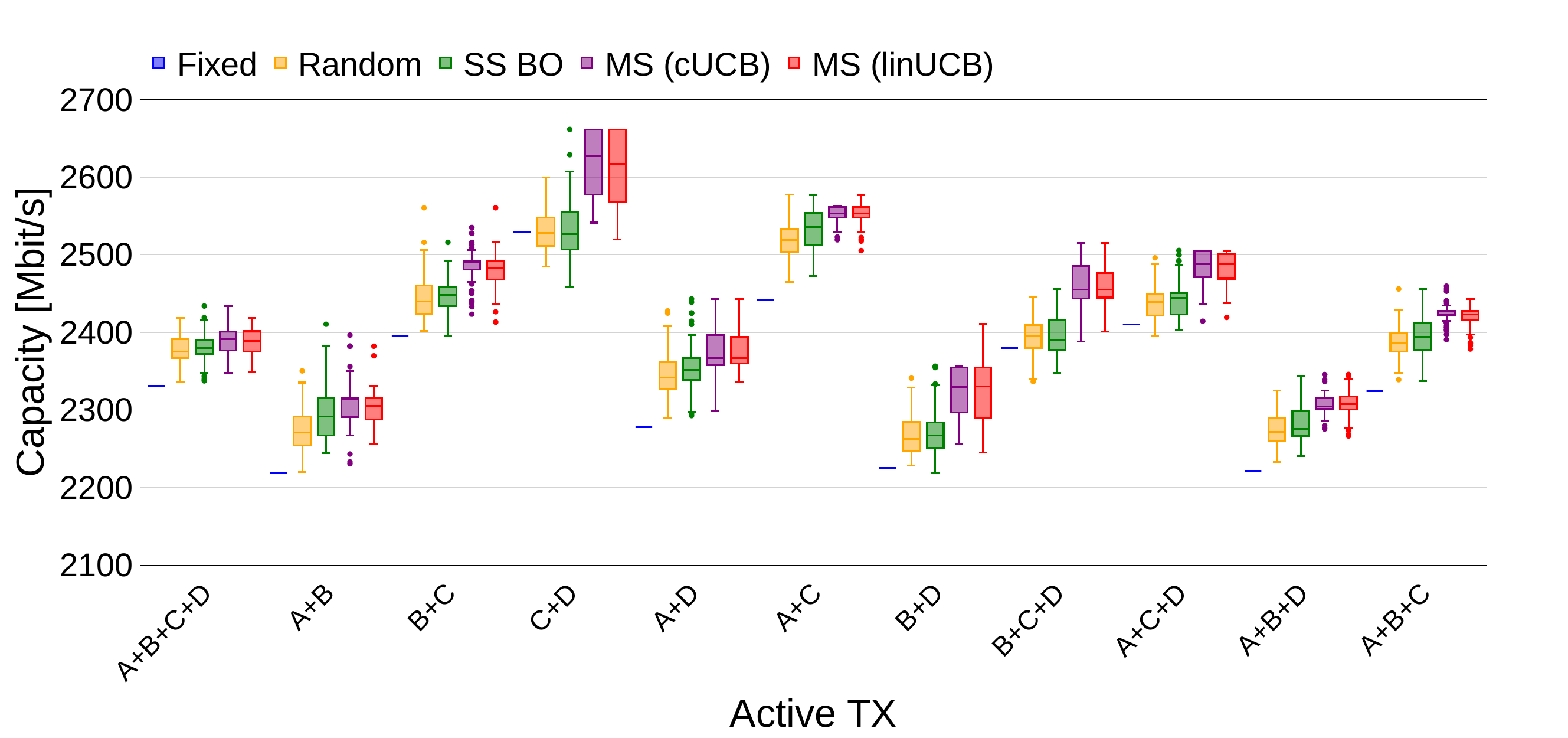}
    \label{fig:real_value_sim_large_merge_seqMA}
  }\\
  \subfloat[\seqMB]{
    \includegraphics[width=\subfigWidth]{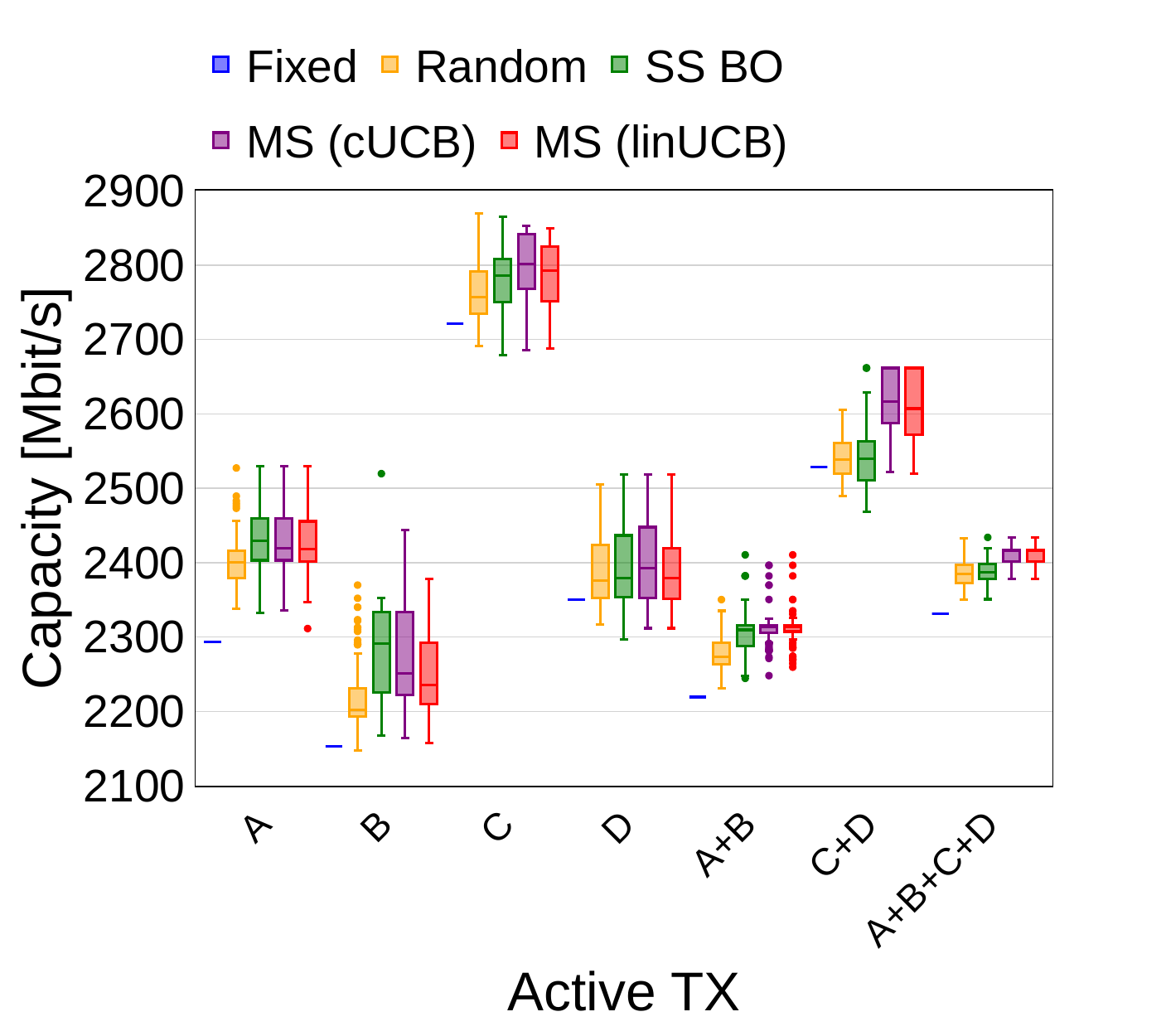}
    \label{fig:real_value_sim_large_merge_seqMB}
  }
  \subfloat[\seqMC]{
    \includegraphics[width=\subfigWidth]{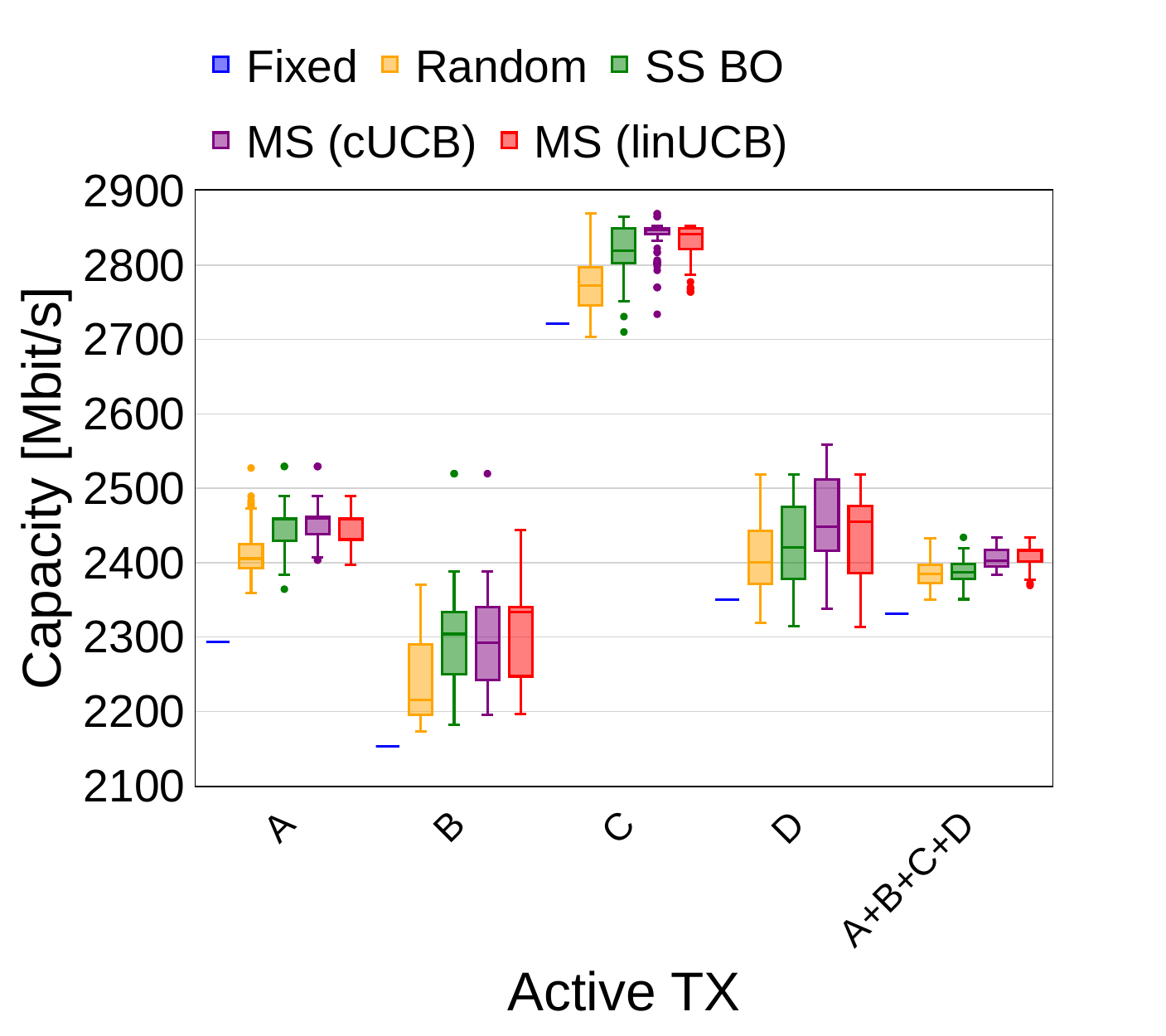}
    \label{fig:real_value_sim_large_merge_seqMC}
  }
  \caption{
    Comparisons of optimization methods for the merged four-link dataset.
    The boxes and lines inside the boxes represent the IQR and median values.
    In (b), (c), and (d), the outliers are plotted as individual points.
    In \seqMA, the capacities of all four \TXpl are observed during the first 20 iterations, allowing the model-sharing frameworks to reuse the learned models when subsets of these links become active.
    In \seqMB and \seqMC, the model-sharing frameworks exhibit a similar trend to that in the two-TX evaluations, achieving more consistent results than state-specific \gls{BO}.
  }
  \label{fig:real_value_sim_large_merge_seqs}
\end{figure}

\newcommand{\seqDef}{
\seqMA: \seqMAdef;
\seqMB: \seqMBdef;
\seqMC: \seqMCdef
}
Fig.~\ref{fig:real_value_sim_large_merge_stats} and Figs.~\ref{fig:real_value_sim_large_merge_seqMA}-\ref{fig:real_value_sim_large_merge_seqMC} show the distributions of the measured channel capacities over all antenna-orientation configurations in the grid and the optimization results for three sequences\footnote{\seqDef} using the merged four-link dataset.
As in the two-link simulations, the model-sharing framework demonstrates greater robustness than state-specific \gls{BO}, while the two acquisition functions achieve comparable performance.
Even in the four-link optimization, where the number of active \glspl{STA} sets become large, the proposed model-sharing framework can efficiently conduct the optimization in unseen active \gls{STA} combinations by reusing the surrogate models learned.

\begin{figure}[!t]
  \centering
  \subfloat[\seqMA\\Active \TX: \IdA, \IdB, \IdC]{
    \includegraphics[width=\subfigWidth]{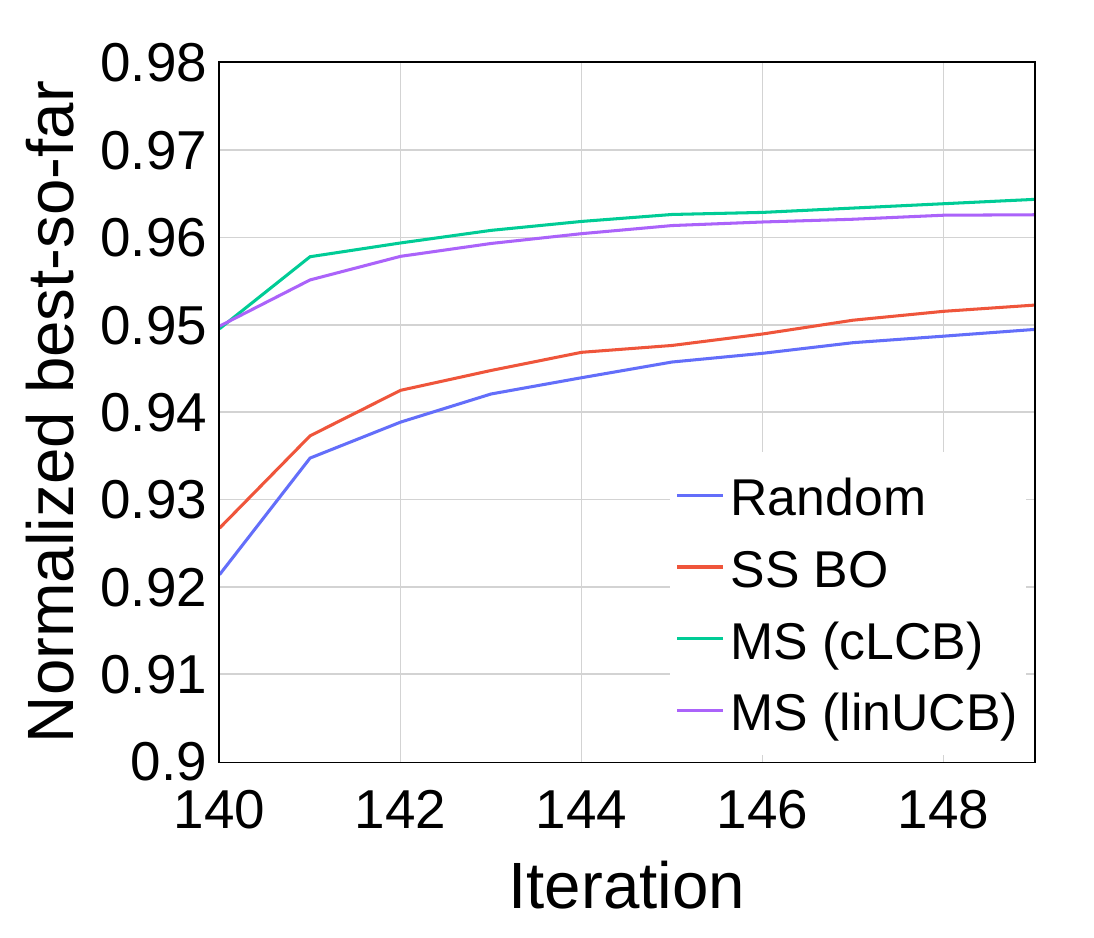}
    \label{fig:real_value_sim_large_merge_trajectory_seqMA}
  }
  \subfloat[\seqMB\\Active \TX: \IdA, \IdB, \IdC, \IdD]{
    \includegraphics[width=\subfigWidth]{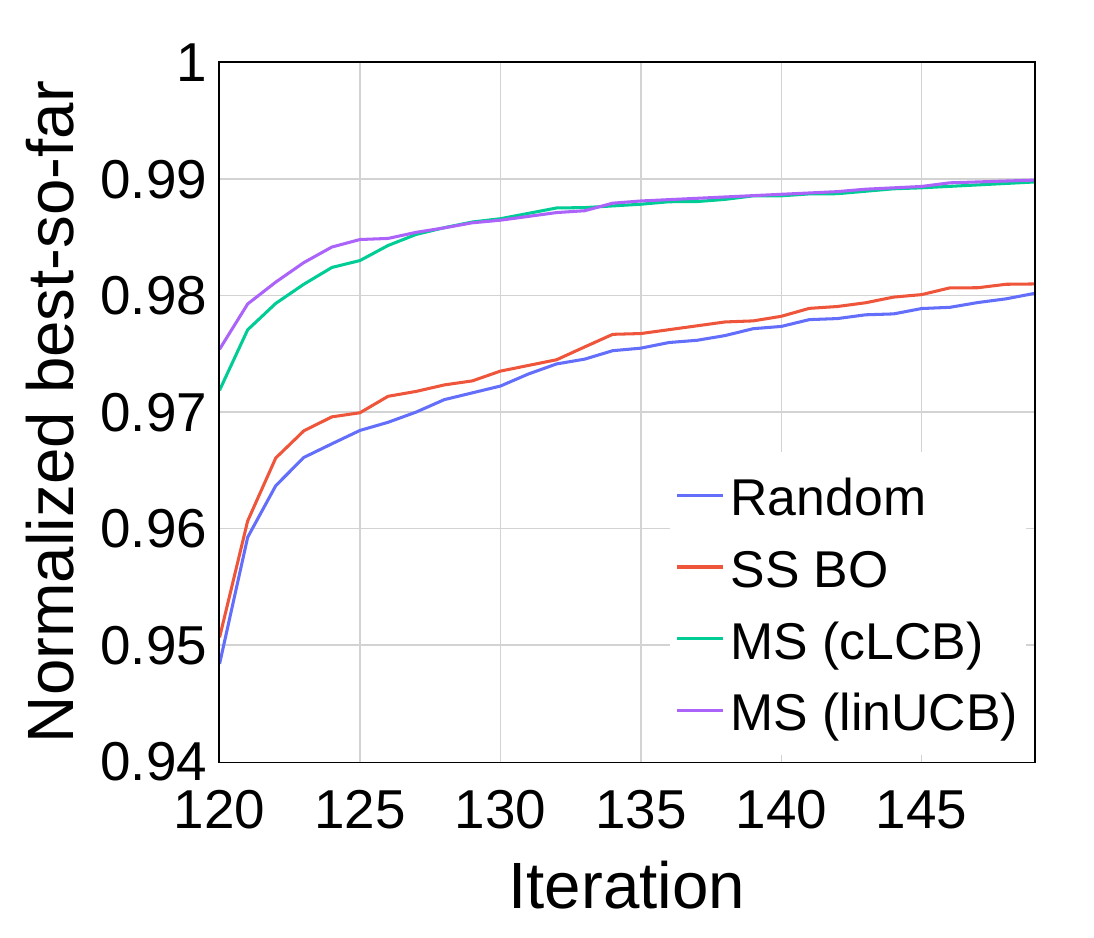}
    \label{fig:real_value_sim_large_merge_trajectory_seqMB}
  }\\
  \subfloat[\seqMC\\Active \TX: \IdA, \IdB, \IdC, \IdD]{
    \includegraphics[width=\subfigWidth]{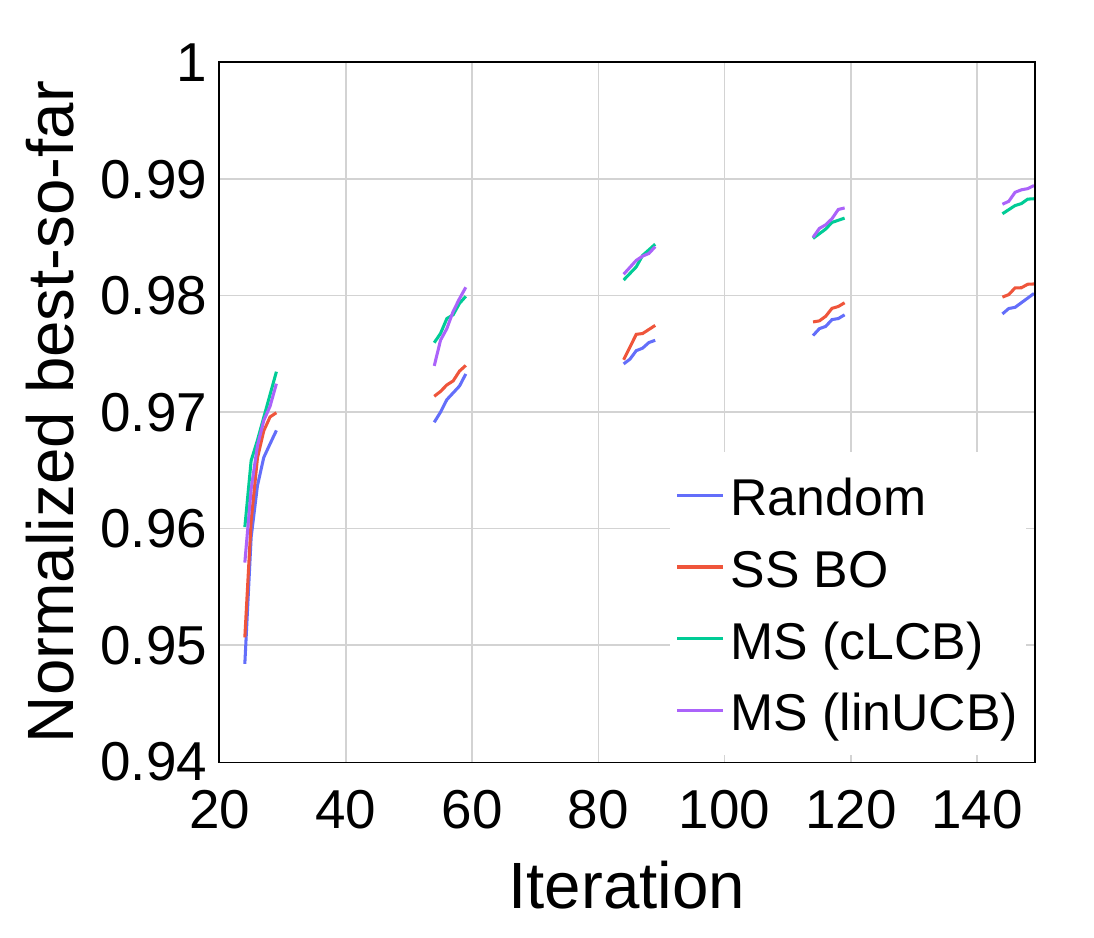}
    \label{fig:real_value_sim_large_merge_trajectory_seqMC}
  }
  \caption{
    Optimization trajectories for the merged four-link dataset.
    The lines represent the mean best-so-far values over 100 trials.
    The metric is the normalized geometric mean of the channel capacity, where the normalization is performed by the per-state optimum.
    The regions corresponding to the specific active \TX sets are extracted from the entire optimization trajectory.
  }
  \label{fig:real_value_sim_large_merge_trajectory}
\end{figure}

The trajectories of the optimization results for the merged four-link dataset are shown in Fig.~\ref{fig:real_value_sim_large_merge_trajectory}.
The lines represent the average values of the best-so-far values among 100 trials, where the metric is the geometric mean of the channel capacity normalized by the per-state optimum.
In \seqMA, the four \TXpl are evaluated at the first 20 iterations, then various combinations of active \TX sets are evaluated.
As shown in Fig.~\ref{fig:real_value_sim_large_merge_trajectory_seqMA}, at the last 10 iterations, where \TX \IdA, \IdB, and \IdC are active, the proposed model-sharing frameworks leverage the trained surrogate models from the previous active \TX sets, and thus achieve higher capacities than random search and state-specific \gls{BO} that start the optimization process from scratch.
Here, the acquisition functions \gls{cUCB} and \gls{propUCB} show comparable performance.
Fig.~\ref{fig:real_value_sim_large_merge_trajectory_seqMB} shows the last 30 iterations of the optimization results for \seqMB, where all four \TXpl are active.
The figure demonstrates the effectiveness of the proposed model-sharing framework as we discussed earlier.
Fig.~\ref{fig:real_value_sim_large_merge_trajectory_seqMC} shows the results for \seqMC, where the active \TX sets are cyclically changed.
Even in frequently changing scenarios, the proposed model-sharing frameworks can efficiently optimize the antenna orientations and achieve higher capacities than the conventional methods.

\subsection{Experimental Evaluation in Real Environment}
\begin{figure}[!t]
  \centering
  \subfloat[Random search]{
    \includegraphics[width=\subfigWidth]{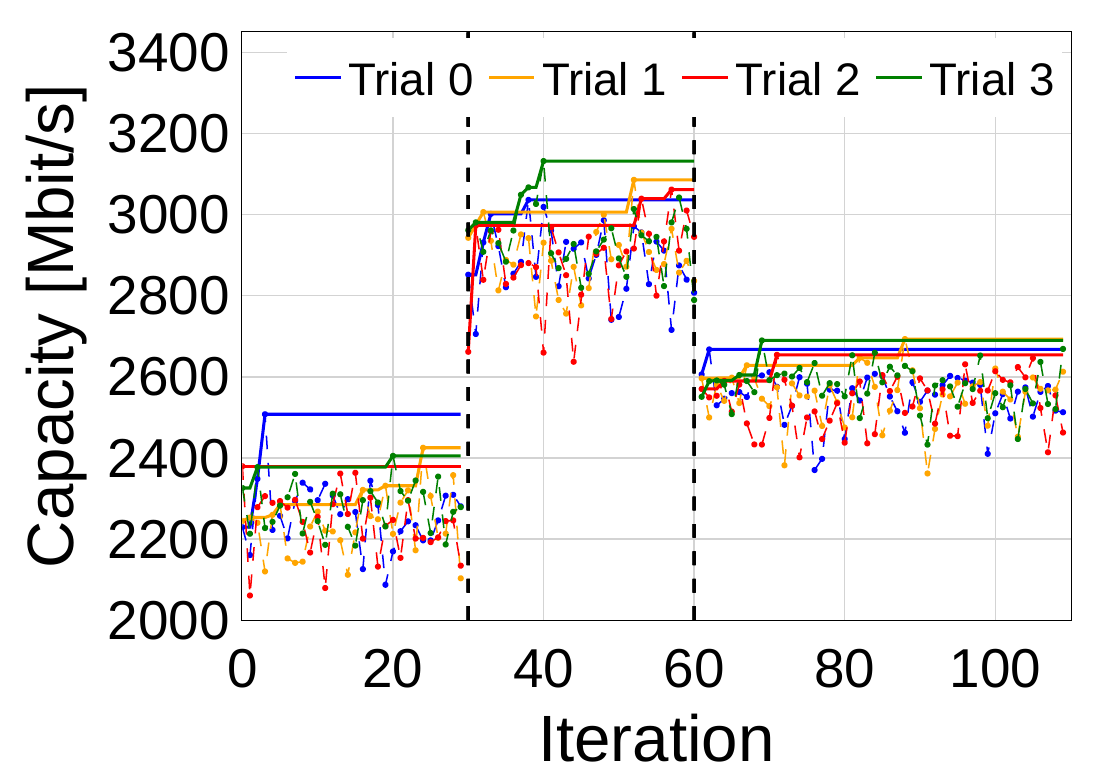}
    \label{fig:experiments_traces_rs}
  }
  \subfloat[State-specific \gls{BO}]{
    \includegraphics[width=\subfigWidth]{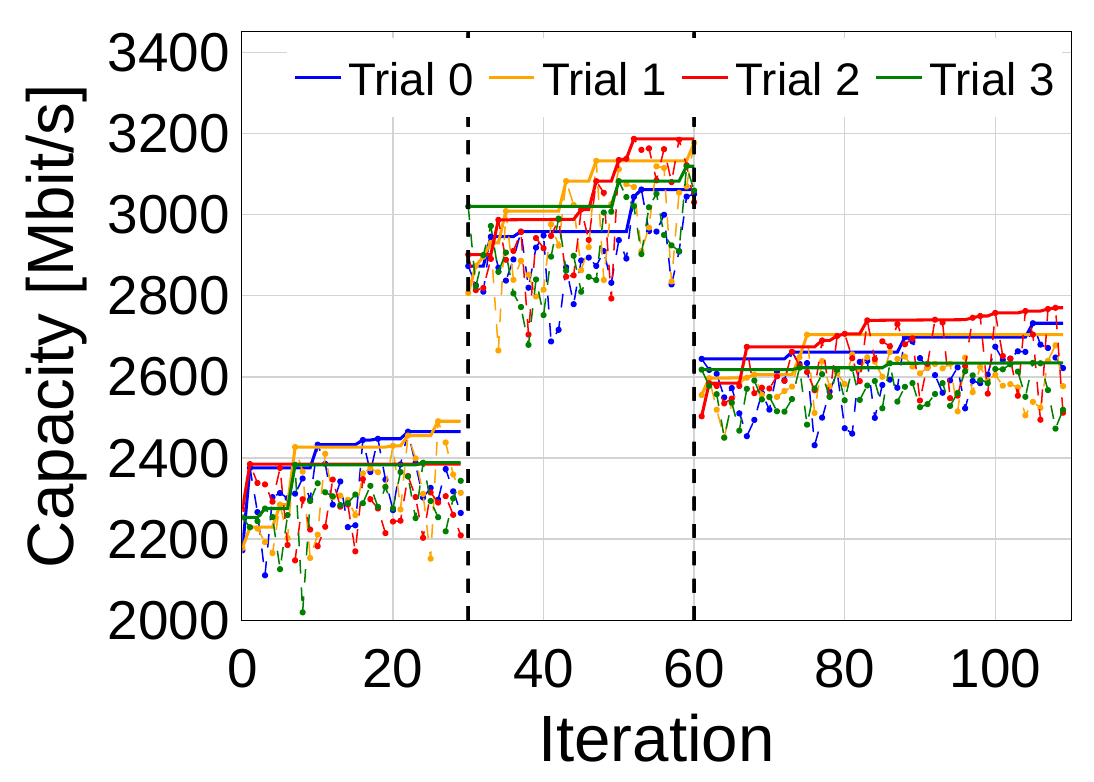}
    \label{fig:experiments_traces_bo}
  }\\
  \subfloat[Model-sharing \gls{BO} with \gls{cUCB}]{
    \includegraphics[width=\subfigWidth]{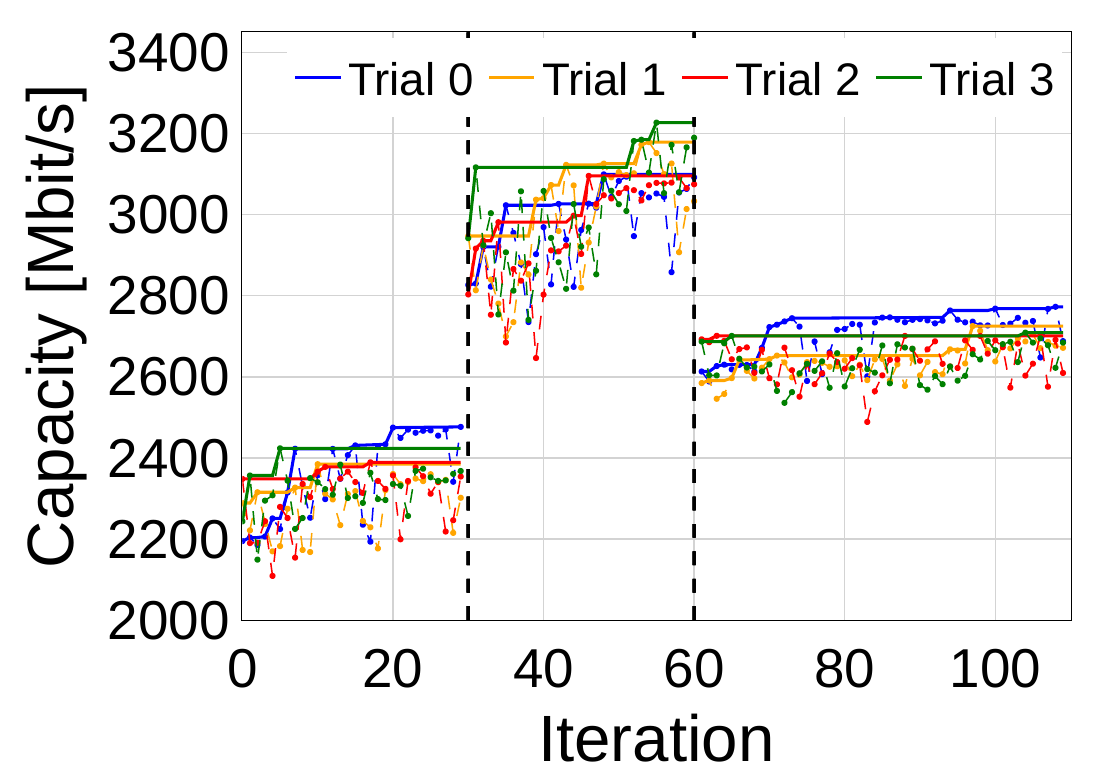}
    \label{fig:experiments_traces_bo_c}
  }
  \subfloat[Model-sharing \gls{BO} with \gls{propUCB}]{
    \includegraphics[width=\subfigWidth]{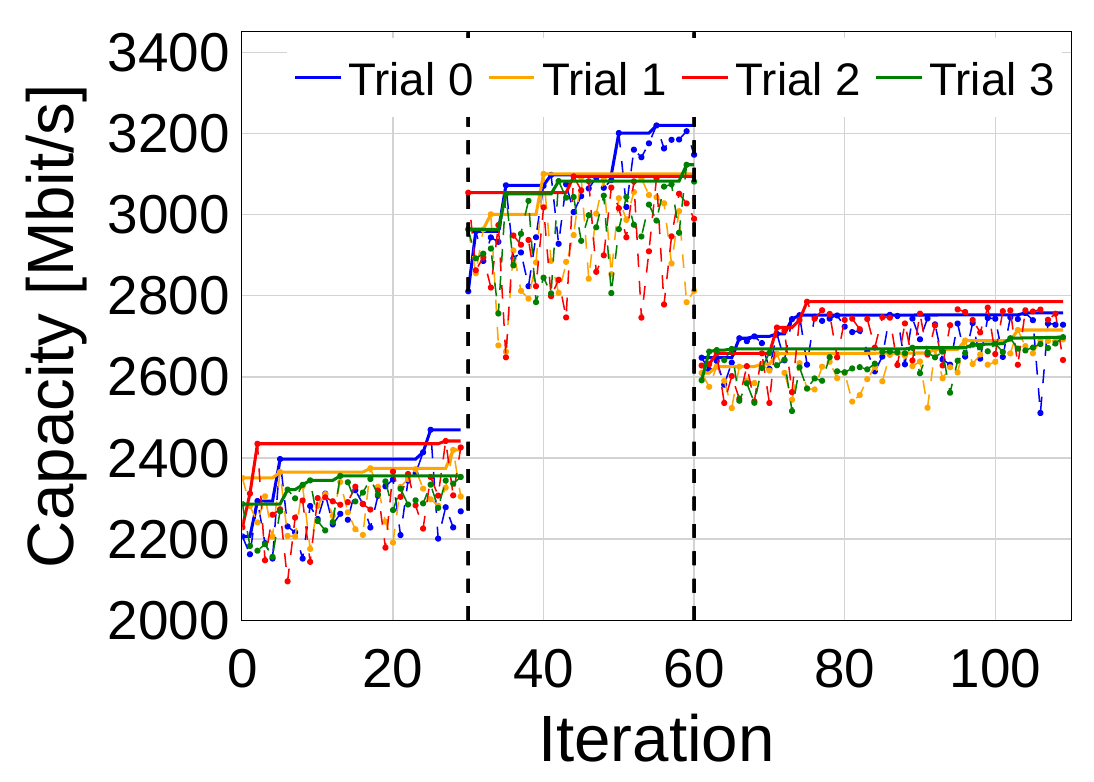}
    \label{fig:experiments_traces_bo_m}
  }
  \caption{
  Optimization trajectories of the channel capacity in the real environment experiments.
  The plotted points and solid lines represent the measured geometric mean of the channel capacity and the best-so-far values, respectively.
  Four trials are conducted for each method, and the results of each trial are shown in different colors.
  The vertical dashed lines indicate the transitions of the active \TXpl.
  First, second, and third sections of the sequence correspond to the single-\TXa, single-\TXb, and both-\TXpl optimization, respectively.
  }
  \label{fig:experiments_traces}
\end{figure}

This subsection presents the evaluation results of the developed antenna orientation optimization device in a real environment.
The experiments were conducted in room \roomNonB using the same devices as in the real-value-based simulations.
Specifically, \TXa was placed at the position labeled \TXa in setting \SetBlock, whereas \TXb was placed at the position labeled \TXa in setting \SetLos, as illustrated in Fig.~\ref{fig:location_roomB}.
\Gls{CSI} measurement frames were injected from the two \TXpl to the proposed device at a period of 100\,msec.
The proposed device calculated the average channel capacity $\AvgCapacity_i(\AngleSet)$ over $\AvgTime=1.0$\,s after waiting 2.0\,s for the antenna actuators to settle.
Active \TXpl were changed according to the sequence \seqRealA.
Active and inactive states were emulated using a software packet filter that blocked measurement frames from inactive \glspl{STA} at the receiver.
The proposed device considered a \TX inactive if no measurement frames were received from it for 1.0\,sec.

We compared the four methods: random search, state-specific \gls{BO}, model-sharing \gls{BO} with \gls{cUCB}, and model-sharing \gls{BO} with \gls{propUCB}.
Each method was evaluated for 4 trials.
The trajectories of the optimization results are shown in Fig.~\ref{fig:experiments_traces}.
The plotted points represent the measured geometric mean of the channel capacity at each iteration, while the solid lines indicate the best-so-far values.
First and second sections of the sequence correspond to the single-\TXa and single-\TXb optimization, respectively, while the third section corresponds to the both-\TXpl optimization.
Random search can occasionally find favorable orientations, so some trials achieve relatively high best-so-far capacity despite many unproductive evaluations.
State-specific \gls{BO} maintains independent models for the three active sets and must learn a new model when both \TXpl become active; many measured capacities in this segment lie between 2,500 and 2,600\,Mbit/s.
In contrast, model-sharing \gls{BO} reuses models trained during the single-\TX segments and more frequently evaluates orientations with capacities between 2,600 and 2,800\,Mbit/s.

\subsection{Discussion and Limitations}
The proposed system optimizes channel capacity estimated from \gls{CSI}, rather than higher-layer throughput.
Actual throughput is also affected by factors such as the \gls{MCS}, contention among multiple \glspl{STA}, and transport- and application-layer protocols.
We assume that the \glspl{STA} share the same \gls{BSS} and transmit separately, and therefore do not explicitly model inter-\gls{STA} interference.
Thus, the optimized metric of the proposed system represents the potential quality of the communication link, which is expected to improve the throughput.
The framework could be applied to higher-layer throughput if the per-link objective can be measured and is sufficiently independent of the active set.
When link throughput depends on interactions among active \glspl{STA}, additional models accounting for these dependencies would be required; this remains future work.

\section{Conclusion} \label{sec:conclusion}
We developed a mechanical antenna-orientation optimization device to improve channel capacity for multiple \glspl{STA} whose activity changes over time.
The proposed system uses a model-sharing \gls{BO} framework to reuse per-link surrogate models across different active-\gls{STA} combinations.
Measurement-based simulations and real-world experiments demonstrated the effectiveness of model sharing relative to state-specific \gls{BO}.

Future work includes extending the framework to mobile \glspl{STA} and moving obstacles, which introduce time-varying propagation conditions beyond the active-set changes considered here.
Such extensions may require dynamic \gls{BO} methods capable of tracking time-varying Gaussian process models.

Another direction is to extend the mechanical antenna device to support both communication and \WiFi sensing.
Applying the device to sensing will require selecting an optimization objective that appropriately reflects sensing performance. 

\section*{Acknowledgment}
The authors used OpenAI's ChatGPT to assist with English translation, proofreading, and stylistic refinement of the manuscript.
The authors reviewed and revised the resulting text and take full responsibility for the manuscript's content.

\bibliographystyle{IEEEtran}
\bibliography{IEEEabrv,myabrv,main}

\begin{IEEEbiography}{Akihito~Taya}
received the B.E.\ degree in electrical and electronic engineering from Kyoto University, Kyoto, Japan in 2011,
and the master and Ph.D.\ degree in Informatics from Kyoto University in 2013 and 2019, respectively.
From 2013 to 2017, he joined Hitachi, Ltd., where he participated in the development of computer clusters.
From 2019 to 2022, he was an assistant professor at Aoyama Gakuin University.
He has been an assistant professor at Institute of Industrial Science, the University of Tokyo since 2022.
He received the IEEE VTS Japan Young Researcher's Encouragement Award in 2012, the IEICE Young Researcher's Award in 2018, and the IEEE CCNC Best Demo Award - Runner Up in 2026.
His current research interests include IoT, distributed machine learning, and Wi-Fi sensing.
He is a member of the IEEE and ACM.
\end{IEEEbiography}

\begin{IEEEbiography}{Yuuki~Nishiyama}
is an Associate Professor at Center for Spatial Information Science (CSIS), the University of Tokyo. He obtained an M.S.\ (2014) in Media and Governance from Keio University and a Ph.D.\ in Media and Governance (2017) from the same institution. He had worked at Keio University in Japan and the University of Oulu in Finland as a post-doctoral researcher, respectively. He began his career at the Institute of Industrial Science, University of Tokyo, as a Research Associate in 2019 and has worked as an Assistant Professor (Lecturer) at CSIS since 2022. Since 2025, he has held his current position. His current research interests include ubiquitous computing systems, mobile and wearable sensing platforms, and human ability augmentation. He is a member of ACM, IEEE, and the Information Processing Society of Japan (IPSJ).
\end{IEEEbiography}

\begin{IEEEbiography}{Kaoru~Sezaki}
is a professor of Center for spatial Science at the University of Tokyo and co-appointed as professor of Institute of Industrial Science, the University of Tokyo. He served as director of Center for Spatial Information Science from 1998 to 2024. He is a steering member of e-Health Technical Committee COMSOC. He has been general chair and TPC Chair of many IEEE international conferences. He also served as Treasurer of IEEE Tokyo Section as well as that of Japan Council from 2003 to 2004. He received B.Eng., M.Eng., and Ph.D.\ degrees from the University of Tokyo, Tokyo, Japan, in 1984, 1986, and 1989, respectively, all in Electrical Engineering. Since 1989, he has been with the University of Tokyo. He was a Visiting Researcher at University of California at San Diego in 1996. His research interests include e-Health, sensor networks, IoT, and urban computing. 
\end{IEEEbiography}

\vfill

\end{document}

%% file: settings.tex
\usepackage{amsmath,amsfonts}
\usepackage{algorithm}
\usepackage{array}
\usepackage{subfig}
\usepackage{textcomp}
\usepackage{stfloats}
\usepackage{url}
\usepackage{verbatim}
\usepackage{graphicx}
\usepackage{cite}
\usepackage{amssymb}
\usepackage{algorithmicx}
\usepackage{algpseudocode}
\usepackage{xcolor}
\usepackage[T1]{fontenc} 
\usepackage[acronym]{glossaries}
\usepackage{mathtools}
\usepackage{textgreek}
\usepackage{bm}
\usepackage{booktabs} 
\usepackage{threeparttable} 
\usepackage{color} 
\usepackage{grffile}
\usepackage{xspace}
\usepackage{tikz}
\usepackage{tikz-3dplot}
\usetikzlibrary{fit,shapes.geometric,decorations.pathreplacing,arrows.meta,calc}
\usepackage{ifthen} 

\algnewcommand{\OnelineIf}[1]{\State\algorithmicif\ #1\ \algorithmicthen}
\algnewcommand{\OnelineForAll}[1]{\State\algorithmicfor\ #1\ \algorithmicdo}

\algblockdefx[ProcedureWoArg]{ProcedureWoArgs}{EndProcedureWoArgs}[1]{\textbf{procedure} #1}{\textbf{end procedure}}
\newcommand{\AlgComment}[1]{\hfill $\triangleright$ #1}

\def\BibTeX{{\rm B\kern-.05em{\sc i\kern-.025em b}\kern-.08em
    T\kern-.1667em\lower.7ex\hbox{E}\kern-.125emX}}

\usepackage{CJKutf8}

\newcommand{\subfigWidthPhoto}{0.2\textwidth}
\newcommand{\subfigWidth}{0.24\textwidth}

%% file: terms.tex
\renewcommand{\[}{\left[}
\renewcommand{\]}{\right]}
\renewcommand{\(}{\left(}
\renewcommand{\)}{\right)}
\DeclareMathOperator*{\argmax}{arg\,max}

\newcommand{\Hermit}{\dagger}
\newcommand{\defeq}{\coloneqq}
\newcommand{\Card}[1]{\lvert#1\rvert}

\newcommand{\Capacity}{C}
\newcommand{\AvgCapacity}{\bar{C}}
\newcommand{\logCap}{l}
\newcommand{\GeoMCapacity}{\hat{C}}
\newcommand{\CapBest}{C^{\mathrm{best}}}
\newcommand{\Fairness}{F}
\newcommand{\FairnessBest}{F^{\mathrm{best}}}

\newcommand{\rssi}{\mathrm{RSSI}}
\newcommand{\NumRxAntenna}{n_{\mathrm{r}}}
\newcommand{\NumTxAntenna}{n_{\mathrm{t}}}
\newcommand{\NumSubcarrier}{K}
\newcommand{\CSI}{H}
\newcommand{\NormCSI}{\tilde{H}}
\newcommand{\SCindex}{k}
\newcommand{\BWperSubCarrier}{\Delta f}
\newcommand{\NoiseDensity}{N_0}
\newcommand{\IdentityMat}{I}
\newcommand{\Boltzman}{k_\mathrm{B}}
\newcommand{\Temperature}{T}
\newcommand{\NoiseFactor}{F}

\newcommand{\Counter}{c}
\newcommand{\InitialPoint}{n_{\mathrm{init}}}

\newcommand{\Stations}{\mathcal{S}}
\newcommand{\RealMat}{\mathbb{R}}
\newcommand{\ComplexMat}{\mathbb{C}}
\newcommand{\AngleSet}{\bm{\theta}}
\newcommand{\AngleSetIdeal}{\AngleSet^*}
\newcommand{\AngleSetBest}{\AngleSet^{\mathrm{best}}}
\newcommand{\StaSet}{S}

\newcommand{\gainCoef}{\alpha}
\newcommand{\WaitTime}{T_\mathrm{wait}}
\newcommand{\AvgTime}{T_\mathrm{avg}}
\newcommand{\threshold}{\gamma_{\mathrm{th}}}
\newcommand{\MaxChangeAccept}{n_{\mathrm{c}}}

\newcommand{\Opt}[1]{\mathcal{O}[#1]}
\newcommand{\OptSuggest}{\Opt{\StaSet}.\mathit{suggest}()}
\newcommand{\OptReportName}{\Opt{\StaSet}.\mathit{report}}
\newcommand{\OptReport}[2]{\OptReportName(#1,#2)}
\newcommand{\OptConv}{\Opt{\StaSet}.\mathit{is\_converged}()}
\newcommand{\OptReset}{\Opt{\StaSet}.\mathit{reset}()}
\newcommand{\OptM}{\tilde{\mathcal{O}}}
\newcommand{\OptMSuggestName}{\OptM.\mathit{suggest}}
\newcommand{\OptMSuggest}[1]{\OptMSuggestName(#1)}
\newcommand{\OptMReportName}{\OptM.\mathit{report}}
\newcommand{\OptMReport}[3]{\OptMReportName(#1,#2,#3)}
\newcommand{\OptMConv}{\OptM.\mathit{is\_converged}()}
\newcommand{\OptMResetName}{\OptM.\mathit{reset}}
\newcommand{\OptMReset}[1]{\OptMResetName(#1)}

\newcommand{\localDS}{\mathcal{D}}
\newcommand{\surrogate}[1]{M_{#1}}
\DeclareMathOperator{\Acq}{UCB}
\newcommand{\ucbCoef}{\kappa}
\newcommand{\meanEst}{\mu}
\newcommand{\stdEst}{\sigma}

\newcommand{\gap}{\delta}

\newcommand{\IdA}{A\xspace}
\newcommand{\IdB}{B\xspace}
\newcommand{\IdC}{C\xspace}
\newcommand{\IdD}{D\xspace}
\newcommand{\TXa}{TX~{\IdA}\xspace}
\newcommand{\TXb}{TX~{\IdB}\xspace}
\newcommand{\TXc}{TX~{\IdC}\xspace}
\newcommand{\TXd}{TX~{\IdD}\xspace}
\newcommand{\TX}{TX\xspace}
\newcommand{\TXpl}{TXs\xspace}
\newcommand{\RX}{RX\xspace}

\newcommand{\roomNon}{layout\xspace}
\newcommand{\roomNonA}{\textsf{a}\xspace}
\newcommand{\roomNonB}{\textsf{b}\xspace}
\newcommand{\roomA}{\roomNon \roomNonA}
\newcommand{\roomB}{\roomNon \roomNonB}
\newcommand{\SetBlock}{I\xspace}
\newcommand{\SetLos}{II\xspace}
\newcommand{\roomBblock}{\roomB-\SetBlock}
\newcommand{\roomBlos}{\roomB-\SetLos}

\newcommand{\seqA}{[\{A\}:50, \{B\}:50, \{A,B\}:50]}
\newcommand{\seqB}{[\{A\}:10, \{B\}:10, \{A,B\}:10]$\times$5}
\newcommand{\seqMA}{sequence \textalpha\xspace}
\newcommand{\seqMB}{sequence \textbeta\xspace}
\newcommand{\seqMC}{sequence \textgamma\xspace}

\newcommand{\seqMAdef}{[\{A,B,C,D\}:20, \{A,B\}:15, \{B,C\}:15, \{C,D\}:15, \{A,D\}:15, \{A,C\}:15, \{B,D\}:15, \{B,C,D\}:10, \{A,C,D\}:10, \{A,B,D\}:10, \{A,B,C\}:10]}
\newcommand{\seqMBdef}{[\{A\}:20, \{B\}:20, \{C\}:20, \{D\}:20, \{A,B\}:20, \{C,D\}:20, \{A,B,C,D\}:30]}
\newcommand{\seqMCdef}{[\{A\}:6, \{B\}:6, \{C\}:6, \{D\}:6, \{A,B,C,D\}:6]$\times$5}

\newcommand{\seqRealA}{[\{A\}:30, \{B\}:30, \{A,B\}:50]}

\newcommand{\etal}{\textit{et al.}\xspace}

\newcommand{\WiFi}{Wi-Fi\xspace}
\newacronym{AI}{AI}{artificial intelligence}
\newacronym{SNR}{SNR}{signal-to-noise ratio}
\newacronym{AP}{AP}{access point}
\newacronym{MIMO}{MIMO}{multiple-input multiple-output}
\newacronym{WLAN}{WLAN}{wireless local area network}
\newacronym{RSSI}{RSSI}{received signal strength indicator}
\newacronym{CSI}{CSI}{channel state information}
\newacronym{MA}{MA}{movable antenna}
\newacronym{RA}{RA}{rotatable antenna}
\newacronym{BO}{BO}{Bayesian optimization}
\newacronym{UCB}{UCB}{upper confidence bound}
\newacronym{cUCB}{cUCB}{composite UCB}
\newacronym{propUCB}{linUCB}{linear UCB}
\newacronym{UAV}{UAV}{unmanned aerial vehicle}
\newacronym{OFDM}{OFDM}{orthogonal frequency-division multiplexing}
\newacronym{STA}{STA}{station}
\newacronym{IoT}{IoT}{Internet of Things}
\newacronym{AIoT}{AIoT}{Artificial IoT}
\newacronym{WebUI}{WebUI}{web user interface}
\newacronym{SDR}{SDR}{software-defined radio}
\newacronym{LOS}{LOS}{line of sight}
\newacronym{ISAC}{ISAC}{integrated sensing and communications}
\newacronym{IQR}{IQR}{interquartile range}
\newacronym{GPR}{GPR}{Gaussian process regression}
\newacronym{LiDAR}{LiDAR}{light detection and ranging}
\newacronym{L-BFGS-B}{L-BFGS-B}{limited-memory Broyden-Fletcher-Goldfarb-Shanno with box constraints}
\newacronym{MCS}{MCS}{modulation and coding scheme}
\newacronym{BSS}{BSS}{basic service set}

%% file: images/sequence.tex
\newcommand{\getdata}[6]{
  \draw[->, thick] (#1, #3) -- node[above, font=\footnotesize, align=center] {#5} (#2, #3);
  \draw[<-, thick, dashed] (#1, #3-#4) -- node[below, font=\footnotesize, align=center] {#6} (#2, #3-#4);
}
\newcommand{\senddata}[4]{
  \draw[->, thick] (#1, #3) -- node[above, font=\footnotesize, align=center] {#4} (#2, #3);
}
\begin{tikzpicture}[>=stealth, font=\sffamily\small]
  \tikzset{
    module/.style={
      draw,
      fill=green!20,
      rounded corners=2pt,
      minimum width=1.0cm,
      minimum height=0.6cm,
      align=center,
    },
    moduleOpt/.style={
      draw=blue,
      fill=blue!20,
      line width=1pt,
      rounded corners=2pt,
      minimum width=1.0cm,
      minimum height=0.6cm,
      align=center,
    },
    storage/.style={
      cylinder,
      draw,
      fill=green!20,
      minimum height=0.8cm,   
      minimum width=1.5cm,    
      cylinder uses custom fill,
      aspect=0.2, 
      rotate=90, 
      align=center,
    },
    sleep/.style={
      draw,
      fill=white,
      minimum width=0.2cm,
      minimum height=#1,
      inner sep=0pt,
      anchor=north,
    }
  }

  \def\txx{0.1}
  \def\csix{1.7}
  \def\bufferx{3.8}
  \def\managerx{5.9}
  \def\optx{7.45}
  \def\actx{8.5}

  \def\mody{0.0}
  \def\suby{-1.2}

  \node[module]    (TX)      at (\txx, \mody) {STA};
  \node[module]    (CsiMod)  at (\csix, \mody) {Capacity\\Calculator};
  \node[storage]   (Buffer)  at (\bufferx, \suby) {\rotatebox{-90}{Buffer}}; 
  \node[module]    (Manager) at (\managerx, \mody) {Optimization\\Manager};
  \node[moduleOpt] (OptMod)  at (\optx, \suby) {Optimizer};
  \node[module]    (ActMod)  at (\actx, \mody) {Motor\\Controller};

  \def\braceYshift{0.6}
  \draw[decorate, decoration={brace, amplitude=6pt}, thick]
    ([yshift=\braceYshift cm]CsiMod.west) -- ([yshift=\braceYshift cm]ActMod.east)
    node[midway, above=6pt] {AP with Mechanical Antenna};

  \def\loopLeftX{\bufferx-0.5}
  \def\loopRightX{\actx+0.5}
  \def\loopTopY{-2.6}
  \def\loopBottomY{-6.9}
  \draw[orange, thin, fill=yellow!10] 
    (\loopLeftX, \loopTopY) rectangle (\loopRightX, \loopBottomY);

  \def\vlineS{-0.3}
  \def\vlineSdouble{-0.4}
  \def\vlineSsub{-1.5}
  \def\vlineE{-7.3}
  \draw[thick] (\txx, \vlineS) -- (\txx, \vlineE);
  \draw[thick] (\csix, \vlineSdouble) -- (\csix, \vlineE);
  \draw[thick] (\bufferx, \vlineSsub) -- (\bufferx, \vlineE);
  \draw[thick] (\managerx, \vlineSdouble) -- (\managerx, \vlineE);
  \draw[thick] (\optx, \vlineSsub) -- (\optx, \vlineE);
  \draw[thick] (\actx, \vlineSdouble) -- (\actx, \vlineE);

  \def\ystart{-0.6}
  \def\ydiff{1.5}
  \def\measureDelay{0.2}
  \foreach \i in {1,...,4} {
    \def\y{\ystart-\i*\ydiff}
    \def\delayedY{\y-\measureDelay}
    \senddata{\txx}{\csix}{\y}{Data}
    \senddata{\csix}{\bufferx}{\delayedY}{Capacity/RSSI}
  }
  \definecolor{darkbrown}{RGB}{101,67,33}
  \node[darkbrown, draw=orange, fill=yellow!50, anchor=north west] at (\loopLeftX, \loopTopY) {optimization loop};
  
  \def\requestDelay{0.1}
  \def\activeCheckY{-4.0}
  \getdata{\managerx}{\bufferx}{\activeCheckY}{\requestDelay}{Check}{active STA}

  \def\suggestionY{\activeCheckY-0.3}
  \def\setAngleY{\suggestionY-1.0}
  \getdata{\managerx}{\optx}{\suggestionY}{\requestDelay}{Request}{suggestion}
  \senddata{\managerx}{\actx}{\setAngleY}{Set angle $\AngleSet$}

  \def\waitStartY{\setAngleY}
  \def\waitDuration{0.8}
  \def\waitEndY{\waitStartY-\waitDuration}
  \node[sleep=\waitDuration cm] (Wait) at (\managerx, \waitStartY) {};
  \node[right, font=\footnotesize, align=left] at ([yshift=2]Wait.east) {Wait $\WaitTime$};

  \def\readCsiY{\waitEndY}
  \def\reportY{\readCsiY-0.3}
  \getdata{\managerx}{\bufferx}{\readCsiY}{\requestDelay}{Request}{average capacity}
  \senddata{\managerx}{\optx}{\reportY}{Report}

\end{tikzpicture}

%% file: images/framework_ss.tex
\begin{tikzpicture}[font=\sffamily\small]
  \tikzset{
    arrow/.style={
      draw,
      color=black,
      line width=0.5pt,
      ->,
      >={Stealth[black,scale=1.3]},
    },
    storage/.style={
      cylinder,
      draw,
      minimum height=0.8,   
      minimum width=3,    
      aspect=0.2, 
      rotate=90, 
      align=center,
    },
    roundrect/.style={
      draw,
      align=center,
      rounded corners=2pt,
    },
  }

  \def\selectorLeft{0.6}
  \def\selectorRight{\selectorLeft+0.5}
  \def\selectorHeight{3.9}
  \def\selectorCenterX{{(\selectorLeft + \selectorRight)/2}} 
  \def\upperLineY{1.4}
  \def\lowerLineY{-1.4}
  \def\optLeft{\selectorRight+0.3}
  \def\optRight{\optLeft+2.8}
  \def\optHeight{1.1}
  \def\dsCenterX{\optLeft+0.5}
  \def\modelCenterX{\optRight-0.9}
  \def\outputLeft{\optRight+0.4}

  \node[roundrect,minimum height=2cm] (INPUT) at (0, 0) {\rotatebox{-90}{Request}};
  \draw[arrow] (INPUT) -- (\selectorLeft, 0);
  \draw[rectangle] (\selectorLeft, \selectorHeight/2) rectangle (\selectorRight, -\selectorHeight/2);
  \node[anchor=center] at (\selectorCenterX, 0) {\rotatebox{-90}{Combination selector}};
  \foreach \y in {\upperLineY, 0, \lowerLineY} {
    \draw[rectangle, fill=blue!5] (\optLeft, \y-\optHeight/2) rectangle (\optRight, \y+\optHeight/2);
    \node[storage] (DS) at (\dsCenterX, \y) {\rotatebox{-90}{DS}};
    \draw[arrow] (\selectorRight, \y) -- (DS.north); 
    \node[roundrect] (GPR) at (\modelCenterX, \y) {GPR $M_i$};
    \draw[arrow] (DS.south) -- (GPR.west);
    \draw[arrow] (GPR.east) -- (\outputLeft, \y);
  }
  \node[roundrect,anchor=west,minimum height=3.5cm] (OUTPUT) at (\outputLeft, 0) {\rotatebox{-90}{Suggestion}};
  \node[color=blue, anchor=south west] at (\optLeft, \upperLineY+0.5) {Optimizers};
\end{tikzpicture}

%% file: images/framework_bo.tex
\begin{tikzpicture}[font=\sffamily\small]
  \tikzset{
    arrow/.style={
      draw,
      color=black,
      line width=0.5pt,
      ->,
      >={Stealth[black,scale=1.3]},
    },
    storage/.style={
      cylinder,
      draw,
      minimum height=0.8,   
      minimum width=3,    
      aspect=0.2, 
      rotate=90, 
      align=center,
    },
    roundrect/.style={
      draw,
      align=center,
      rounded corners=2pt,
    },
  }

  \def\upperLineY{0.7}
  \def\lowerLineY{-0.7}
  \def\optLeft{0.6}
  \def\optRight{\optLeft+5.2}
  \def\optHeight{3.4}
  \def\selectorLeft{\optLeft+0.2}
  \def\selectorRight{\selectorLeft+0.55}
  \def\selectorHeight{2.5}
  \def\selectorCenterX{{(\selectorLeft + \selectorRight)/2}} 
  \def\dsCenterX{\selectorRight+0.7}
  \def\modelCenterX{\dsCenterX+1.5}
  \def\sumCenterX{\modelCenterX+1.7}
  \def\outputX{\optRight+0.8}

  \node[roundrect,minimum height=2cm] (INPUT) at (0, 0) {\rotatebox{-90}{Request}};
  \draw[rectangle, fill=blue!5] (\optLeft, \optHeight/2) rectangle (\optRight, -\optHeight/2);
  \draw[arrow] (INPUT) -- (\selectorLeft, 0);
  \draw[rectangle] (\selectorLeft, \selectorHeight/2) rectangle (\selectorRight, -\selectorHeight/2);
  \node[anchor=center] at (\selectorCenterX, 0) {\rotatebox{-90}{STA selector}};
  \node[draw,circle] (SUM) at (\sumCenterX, 0) {$\sum$};
  \foreach \y in {\upperLineY, \lowerLineY} {
    \node[storage] (DS) at (\dsCenterX, \y) {\rotatebox{-90}{DS}};
    \draw[arrow] (\selectorRight, \y) -- (DS.north); 
    \node[roundrect] (GPR) at (\modelCenterX, \y) {GPR $M_i$};
    \draw[arrow] (DS.south) -- (GPR.west);
    \draw[arrow] (GPR.east) -- (SUM);
  }
  \node[roundrect,minimum height=2.5cm] (OUTPUT) at (\outputX, 0) {\rotatebox{-90}{Suggestion}};
  \draw[arrow] (SUM.east) -- (OUTPUT.west);
  \node[color=blue, anchor=south west] at (\optLeft, \upperLineY+1.0) {Optimizer};
\end{tikzpicture}

%% file: images/angle_def.tex
\tdplotsetmaincoords{65}{150}
\begin{tikzpicture}[scale=1.3, tdplot_main_coords]
  \node at (0, -0.5, -0.0) {\includegraphics[width=\subfigWidth,trim=0 0 120 0,clip]{images/device/top.jpg}};

  \tikzset{
    arc/.style={
      double=green,
      double distance=0.5pt,
      line width=0.3pt,
      color=black,
      draw,
    },
    coordinate/.style={
      double=green,
      double distance=1pt,
      line width=0.5pt,
      color=black,
      draw,
      dashed,
    },
    arc arrow/.style={
      double=green,
      double distance=1pt,
      line width=0.5pt,
      color=black,
      draw,
      ->,
      >={Stealth[black,scale=1.3,fill=green]},
    },
    label/.style={
      color=black,
      fill=white,
      font=\small,
      inner sep=1pt,
    },
  }


  \def\leftAnt{-1.1}
  \def\rightAnt{1.1}
  \def\antLen{1.5}
  \def\angleR{0.5}
  \def\leftAnchor{(\antLen*cos{30}, \leftAnt+\antLen*sin{30}, 0)}
  \def\rightAnchor{(\antLen*cos{30}, \rightAnt-\antLen*sin{30}, 0)}

  \draw[coordinate] (0, \leftAnt, 0) -- (0, \leftAnt, \antLen);
  \draw[coordinate] (0, \leftAnt, 0) -- (\antLen, \leftAnt, 0);
  \draw[coordinate] (0, \leftAnt, 0) -- (0, \leftAnt-\antLen, 0);
  \draw[coordinate] (0, \leftAnt, 0) -- \leftAnchor;
  \draw[coordinate] (0, \rightAnt, 0) -- (0, \rightAnt, \antLen);
  \draw[coordinate] (0, \rightAnt, 0) -- (\antLen, \rightAnt, 0);
  \draw[coordinate] (0, \rightAnt, 0) -- (0, \rightAnt+\antLen, 0);
  \draw[coordinate] (0, \rightAnt, 0) -- \rightAnchor;

  \draw[arc arrow] \leftAnchor arc (30:-90:\antLen);
  \draw[arc arrow] \rightAnchor arc (-30:90:\antLen);

  \begin{scope}[shift={(0, \leftAnt, 0)}, rotate around x=90]
    \draw[arc] (0, \antLen, 0) arc (90:0:\antLen);
    \draw[arc arrow] (0, \angleR, 0) arc (90:0:\angleR);
  \end{scope}
  \begin{scope}[shift={(0, \rightAnt, 0)}, rotate around x=90]
    \draw[arc] (0, \antLen, 0) arc (90:0:\antLen);
    \draw[arc arrow] (0, \angleR, 0) arc (90:0:\angleR);
  \end{scope}

  \node[label] at (0, \leftAnt, \angleR) [xshift=1.1em, yshift=0.1em] {0\textdegree};
  \node[label] at (0, \rightAnt, \angleR) [xshift=1.1em, yshift=0.1em] {0\textdegree};
  \node[label] at (\angleR, \leftAnt, 0) [xshift=-1.2em, yshift=0.4em] {90\textdegree};
  \node[label] at (\angleR, \rightAnt, 0) [xshift=0.1em, yshift=-0.7em] {90\textdegree};
  \node[label] at (0, \leftAnt, 0) [xshift=-1.4em, yshift=1.4em] {$\theta_0$};
  \node[label] at (0, \rightAnt, 0) [xshift=-1.4em, yshift=1.4em] {$\theta_1$};

  \node[label] at \leftAnchor [xshift=-0.8em, yshift=-0.6em] {60\textdegree};
  \node[label] at \rightAnchor [xshift=-0.8em, yshift=-0.6em] {60\textdegree};
  \node[label] at (0, \leftAnt-\antLen, 0) [xshift=0.4em, yshift=0.8em] {180\textdegree};
  \node[label] at (0, \rightAnt+\antLen, 0) [xshift=1.0em, yshift=0.0em] {180\textdegree};
  \node[label] at (\antLen, \leftAnt, 0) [xshift=-1.0em, yshift=-0.4em] {$\varphi_0$};
  \node[label] at (\antLen, \rightAnt, 0) [xshift=0.4em, yshift=-1.2em] {$\varphi_1$};

  \node[label, anchor=west] at (0, \leftAnt, \antLen) [xshift=0.4em, yshift=-0.1em] {Antenna 0};
  \node[label, anchor=west] at (0, \rightAnt, \antLen) [xshift=0.6em, yshift=0.1em] {Antenna 1};
\end{tikzpicture}

%% file: images/location_small.tex
\begin{tikzpicture}
  \tikzset{
    arrow/.style={
      thin,
      dashed,
      {Stealth[length=7,width=6]}-{Stealth[length=7,width=6]}
    },
    eventLabel/.style={
      font=\footnotesize\sffamily,
      color=#1,
      align=left
    },
    desk/.style={
      draw,
      fill=gray!20,
      thin,
    },
  }
  \def\left{0.4}
  \def\right{7.0}
  \def\top{3.6}
  \def\bottom{-0.1}

  \newcommand{\device}[4]{
    \node[rectangle, draw, thin, minimum width=19pt, minimum height=11pt] #1 at #2 {};
    \node[font=\scriptsize \bfseries, color=#3] at #2 {\textsf{#4}};
  }
  \newcommand{\ladder}[2]{
    \def\legwidth{0.08}
    \def\delta{0.07} 
    \def\left{#1 + 0.06}
    \def\right{#1 + 0.9}
    \def\top{#2 + 1.0}
    \def\second{#2 + 0.70}
    \def\third{#2 + 0.4}
    \def\bottom{#2 + 0.06}

    \draw[line width=4pt] (\left, \bottom)  -- (\left + \legwidth, \top);
    \draw[line width=4pt] (\right, \bottom) -- (\right - \legwidth, \top);

    \foreach \y in {\top, \second, \third} {
      \draw[line width=4pt] (\left + \legwidth - \delta, \y) -- (\right - \legwidth + \delta, \y);
    }
  }

  \device{(RX)}{(1.05, 3.25)}{black}{\RX};
  \ladder{0.58}{1.9}


  \draw[desk] (3.2, 0.0) rectangle (4.9, 2.9);
  \node[inner sep=2] at (4.9, 0.0) [anchor=south east] {Desk};

  \def\RxCenter{1.05};
  \def\TxACenter{3.8};
  \def\TxAHeight{2.5};
  \def\TxBCenter{2.4};
  \def\TxBNextHeight{0.9};
  \def\LineBottomA{0.6}
  \def\LineBottomB{0.6}
  \def\ArrowAheight{1.0};
  \def\ArrowBheight{1.4};

  \device{(TXa)}{(\TxACenter, \TxAHeight)}{black}{\TXa};
  \device{(TXb)}{(\TxACenter, \TxBNextHeight)}{black}{\TXb};

  \draw[draw, thin] (RX.south) -- (\RxCenter, \LineBottomA);
  \draw[arrow] (\RxCenter, \ArrowBheight) -- (\TxACenter, \ArrowBheight) node[midway, below, inner sep=2] {2.2\,m};
  \draw[arrow] (TXa.south) -- (TXb.north) node[midway, right] {1.3\,m};
\end{tikzpicture}

%% file: images/location.tex
\begin{tikzpicture}[scale=1.3]
  \tikzset{
    arrow/.style={
      thin,
      dashed,
      {Stealth[length=7,width=6]}-{Stealth[length=7,width=6]}
    },
    desk/.style={
      draw,
      fill=gray!20,
      thin,
    },
  }

  \def\deskLeft{0}
  \def\deskRight{1.8}
  \def\deskBottom{1.98}
  \def\deskTop{4.5}
  \def\deskWidth{1.2}
  \def\deskHeight{0.4}
  \def\cartWidth{0.45}
  \def\cartHeight{0.7}
  \def\boardWidth{0.30}
  \def\boardHeight{0.05}

  \def\ladderLeft{0.0}
  \def\ladderBottom{\deskTop+0.36}

  \def\leftLine{\deskLeft+0.3}
  \def\rightLine{\deskRight+0.28}
  \def\upperLine{\deskTop+0.25}
  \def\lowerLine{\deskBottom+0.25}
  \def\rxLine{\deskTop+1.06}

  \def\cartLeft{\leftLine-\cartWidth/2}
  \def\cartBottom{\deskBottom+\deskHeight+0.05}
  \def\boardLeft{\leftLine-\boardWidth/2}
  \def\boardBottom{\cartBottom+0.05}
  \coordinate (boardCoord) at (\leftLine, \boardBottom+\boardHeight) {};
  \coordinate (boardLabel) at (\boardLeft+\boardWidth/2+0.3, \boardBottom+0.40) {};
  \coordinate (ladderLabel) at (\ladderLeft+0.6, \ladderBottom+0.35) {};
  \coordinate (layoutBlockLabelCoord) at (\deskLeft+1.9, \deskBottom+1.6) {};
  \coordinate (layoutLosLabelCoord) at (\deskLeft+0.4, \deskBottom+1.8) {};

  \definecolor{brown}{RGB}{181,68,39}

  \newcommand{\device}[4]{
    \node[rectangle, draw, thin, minimum width=19pt, minimum height=11pt] #1 at #2 {};
    \node[font=\scriptsize \bfseries, color=#3] at #2 {\textsf{#4}};
  }

  \newcommand{\ladder}[2]{
    \def\legwidth{0.05}
    \def\delta{0.04} 
    \def\left{#1 + 0.03}
    \def\right{#1 + 0.57}
    \def\top{#2 + 0.5}
    \def\second{#2 + 0.35}
    \def\third{#2 + 0.2}
    \def\bottom{#2 + 0.03}

    \draw[line width=3pt] (\left, \bottom)  -- (\left + \legwidth, \top);
    \draw[line width=3pt] (\right, \bottom) -- (\right - \legwidth, \top);

    \foreach \y in {\top, \second, \third} {
      \draw[line width=3pt] (\left + \legwidth - \delta, \y) -- (\right - \legwidth + \delta, \y);
    }
  }

  \newcommand{\labeledVArrow}[3]{
    \draw[arrow] #1 -- node[midway, left, font=\footnotesize] {\rotatebox{90}{#3}} #2;
  }
  \newcommand{\labeledHArrow}[3]{
    \draw[arrow] #1 -- node[midway, above, font=\footnotesize] {#3} #2;
  }

  \draw[desk] (\deskLeft, \deskBottom) rectangle (\deskLeft+\deskWidth, \deskBottom+\deskHeight);
  \draw[desk] (\deskRight, \deskBottom) rectangle (\deskRight+\deskWidth, \deskBottom+\deskHeight);
  \draw[desk] (\deskLeft, \deskTop) rectangle (\deskLeft+\deskWidth, \deskTop+\deskHeight);
  \draw[desk] (\deskRight, \deskTop) rectangle (\deskRight+\deskWidth, \deskTop+\deskHeight);
  \node[anchor=south east,font=\footnotesize] at (\deskRight+\deskWidth, \deskBottom) {Desk};

  \device{(RX)}{(\leftLine, \rxLine)}{black}{\RX}
  \ladder{\ladderLeft}{\ladderBottom}
  \node[anchor=west, inner sep=2pt, font=\footnotesize] at (ladderLabel) {Ladder with height of 1.7\,m};

  \draw[draw, fill=blue!10] (\cartLeft, \cartBottom) rectangle (\cartLeft+\cartWidth, \cartBottom+\cartHeight);
  \draw[draw, fill=gray] (\boardLeft, \boardBottom) rectangle (\boardLeft+\boardWidth, \boardBottom+\boardHeight);
  \node[anchor=west, inner sep=2pt, font=\footnotesize] at (boardLabel) {Aluminum plate on a cart};
  \draw (boardCoord) -- (boardLabel);

  \device{(TXA)}{(\rightLine, \upperLine)}{blue}{\TXa}; 
  \device{(TXB)}{(\leftLine, \lowerLine)}{blue}{\TXb}; 

  \device{(TXC)}{(\leftLine, \upperLine)}{brown}{\TXa}; 
  \device{(TXD)}{(\rightLine, \lowerLine)}{brown}{\TXb}; 


  \def\arrowAX{\deskLeft-0.1}
  \def\arrowBX{\deskLeft-0.5}
  \def\arrowCY{\rxLine+0.4}
  \def\lineStartX{\arrowBX-0.15}
  \def\lineMiddleX{\arrowAX-0.15}
  \def\lineEndY{\arrowCY+0.15}

  \draw[thin] (RX) -- (\lineStartX, \rxLine);
  \draw[thin] (TXC) -- (\lineMiddleX, \upperLine);
  \draw[thin] (TXB) -- (\lineStartX, \lowerLine);

  \labeledVArrow{(\arrowAX, \rxLine)}{(\arrowAX, \upperLine)}{0.7\,m}
  \labeledVArrow{(\arrowAX, \rxLine)}{(\arrowAX, \upperLine)}{0.7\,m}
  \labeledVArrow{(\arrowBX, \rxLine)}{(\arrowBX, \lowerLine)}{4.6\,m}

  \draw[thin] (RX) -- (\leftLine, \lineEndY);
  \draw[thin] (TXA) -- (\rightLine, \lineEndY);

  \labeledHArrow{(\leftLine, \arrowCY)}{(\rightLine, \arrowCY)}{2.2\,m}

  \node[color=blue, font=\footnotesize, anchor=center, inner sep=2pt] (layoutBlockLabel) at (layoutBlockLabelCoord) {\roomBblock};
  \node[color=brown, font=\footnotesize, anchor=center, inner sep=2pt] (layoutLosLabel) at (layoutLosLabelCoord) {\roomBlos};
  \draw[color=blue, thin] (TXA) -- (layoutBlockLabel);
  \draw[color=blue, thin] (TXB) -- (layoutBlockLabel);
  \draw[color=brown, thin] (TXC) -- (layoutLosLabel);
  \draw[color=brown, thin] (TXD) -- (layoutLosLabel);

\end{tikzpicture}